\documentclass[pre,tighten,showpacs]{revtex4}

\usepackage{amssymb,amsmath}

\usepackage{graphicx}% Include figure files

\begin{document}
\title{Transport coefficients for an inelastic gas around uniform shear flow: Linear
stability analysis}
\author{Vicente Garz\'{o}}
\affiliation{Departamento de F\'{\i}sica, Universidad de Extremadura, E-06071
Badajoz, Spain}
\email[E-mail: ]{vicenteg@unex.es}

\begin{abstract}
The inelastic Boltzmann equation for a granular gas is applied to
spatially inhomogeneous states close to the uniform shear flow. A
normal solution is obtained via a Chapman-Enskog-like expansion
around a local shear flow distribution. The heat and momentum
fluxes are determined to first order in the deviations of the
hydrodynamic field gradients from their values in the reference
state. The corresponding transport coefficients are determined
from a set of coupled linear integral equations which are
approximately solved by using a kinetic model of the Boltzmann
equation. The main new ingredient in this expansion is that the
reference state $f^{(0)}$ (zeroth-order approximation) retains all
the hydrodynamic orders in the shear rate. In addition, since the
collisional cooling cannot be compensated locally for viscous
heating, the distribution $f^{(0)}$ depends on time through its
dependence on temperature. This means that in general, for a given
degree of inelasticity, the complete nonlinear dependence of the
transport coefficients on the shear rate requires the analysis of
the {\em unsteady} hydrodynamic behavior. To simplify the
analysis, the steady state conditions have been considered here in
order to perform a linear stability analysis of the hydrodynamic
equations with respect to the uniform shear flow state. Conditions
for instabilities at long wavelengths are identified and
discussed.
\end{abstract}

\draft
\pacs{ 05.20.Dd, 45.70.Mg, 51.10.+y, 47.50.+d}
\date{\today}
\maketitle

\section{Introduction}

The understanding of granular systems still remains a topic of
interest and controversy. Under rapid flow conditions, they can be
modeled as a fluid of hard spheres dissipating part of their
kinetic energy during collisions. In the simplest model, the
grains are taken to be smooth so that the inelasticity of
collisions is characterized through a constant coefficient of
normal restitution $\alpha \leq 1$. Energy dissipation has
profound consequences on the behavior of these systems since they
exhibit a rich phenomenology with many qualitative differences
with respect to molecular systems. In particular, the absence of
energy conservation yields subtle modifications of the
conventional Navier-Stokes equations for states with small
gradients of the hydrodynamic fields. The dependence of the
corresponding transport coefficients on dissipation may be
determined from the Boltzmann kinetic equation conveniently
modified to account for inelastic binary collisions
\cite{GS95,BDS97}. The idea is to extend the Chapman-Enskog method
\cite{CC70} to the inelastic case by expanding the velocity
distribution function around the local version of the homogeneous
cooling state, namely, a homogeneous state whose dependence on
time occurs only through the temperature. In the first order of
the expansion, explicit expressions for the transport coefficients
as functions of the coefficient of restitution have been obtained
in the case of a single gas \cite{single} as well as for granular
mixtures \cite{mixture}, showing good agreement the analytical
results with those obtained from Monte Carlo simulations
\cite{DSMC}.

Although the Chapman-Enskog method can be in principle applied to
get higher orders in the gradients (Burnett and super-Burnett
corrections,$\ldots$), it is extremely difficult to evaluate those
terms especially for inelastic systems. In addition, questions
about its convergence remains still open \cite{GS03}. This gives
rise to the search for alternative approaches to characterize
transport for strongly inhomogeneous situations (i.e., beyond the
Navier-Stokes limit). One possibility is to expand in small
gradients around a more relevant reference state than the (local)
homogenous cooling state. For example,
consider states near a shearing reference steady state such as the
so-called uniform (simple) shear flow (USF) \cite{C90}.
Such an application of the
Chapman-Enskog method to a nonequilibrium state requires some
care as recently discussed in Ref.\ \cite{L05}.  The USF
state is probably the simplest flow problem since the only nonzero
hydrodynamic gradient is $\partial u_x/\partial y\equiv
a=\text{const}$, where ${\bf u}$ is the flow velocity and $a$ is
the constant shear rate. Due to its simplicity, this state has
been widely used in the past both for elastic \cite{GS03} and
inelastic gases \cite{C90} to shed light on the complexities
associated with the {\em nonlinear} response of the system to the
action of strong shearing. However, the nature of this state for
granular systems is different from that of the elastic fluids
since the source of energy due to the macroscopic imposed shear
field drives the granular system into rapid flow and a steady
state is achieved when the amount of energy supplied by shearing
work is balanced by the lost one due to the inelastic collisions
between the particles. As a consequence, in the steady state the
reduced shear rate $a^*\propto a/\sqrt{T}$ (which is the relevant
nonequilibrium parameter of the problem) is not an independent
quantity but becomes a function of the coefficient of restitution
$\alpha$. This means that the {\em quasielastic} limit ($\alpha\to
1$) naturally implies the limit of {\em small} shear rates
($a^*\ll 1$) and vice versa. The study of the rheological
properties of the USF state has received a great deal of attention
in recent years in the case of monocomponent
\cite{LSJC84,JR88,C89,HS92,
SK94,LB94,GT96,SGN96,BRM97,CR98,MGSB99} and multicomponent systems
\cite{CH02,MG02,G02,L04,AL03, MGAL05}.

The aim of this paper is to determine the heat and momentum fluxes
of a gas of inelastic hard spheres under simple shear flow in the
framework of the Boltzmann equation. The physical situation is
such that the gas is in a state that deviates from the simple
shear flow by {\em small} spatial gradients. The starting point of
this study is a recent approximate solution of the Boltzmann
equation which is based on Grad's method \cite{MG02,G02,SGD04}. In
spite of this approach, the relevant transport properties obtained
from this solution compare quite well with Monte Carlo simulations
even for strong dissipation \cite{BRM97,MG02,AS05}, showing again
the reliability of Grad's approximation to compute the lowest
velocity moments of the velocity distribution function. Since the
system is slightly perturbed form the USF, the Boltzmann equation
is solved by applying the Chapman-Enskog method around the (local)
shear flow state rather than the (local) homogeneous cooling
state. This is the main feature of this expansion since the
reference state is not restricted to small values of the shear
rate. One important point is that, for general small deviations
from the shear flow state, the zeroth-order distribution
is not a stationary distribution since the
collisional cooling cannot be compensated {\em locally} for
viscous heating. This fact gives rise to new conceptual and
practical difficulties not present in the previous analysis made
for elastic gases to describe transport in thermostatted shear
flow states \cite{LD97}. Due to the difficulties involved in this
expansion, here general results will be restricted to particular
perturbations for which steady state conditions apply. In the
first order of the expansion, the generalized transport
coefficients are given in terms of the solutions of linear
integral equations. To get explicit expressions for these
coefficients, one needs to know the fourth-degree moments of USF.
This requires to consider higher-order terms in Grad's
approximation for the reference distribution function, which is
quite an intricate problem. In order to overcome such difficulty,
here I have used a convenient kinetic model \cite{BDS99} that
preserves the essential properties of the inelastic Boltzmann
equation but admits more practical analysis. The mathematical and
physical basis for this model as a good representation of the
Boltzmann equation has been discussed in Ref.\ \cite{BDS99}. In
particular, it is worth noting that the results derived from this
model coincides with those given from the Boltzmann equation at
the level of the rheological properties \cite{BRM97,BDS99}.
Furthermore, recent computer simulation results \cite{AS05} have
also shown good agreement between the kinetic model and the
Boltzmann equation for the fourth-degree moments, covering this
agreement a wide range of values of dissipation (say, for
instance, $\alpha \gtrsim 0.5$). This good agreement extends that
previously demonstrated for Couette flow in dilute gases
\cite{TTMGSD01} and for USF in dense systems \cite{MGSB99} and
shows the reliability of the kinetic model to capture the main
trends of the Boltzmann equation, especially those related to
transport properties.

The knowledge of the above generalized transport coefficients
allows one to determine the hydrodynamic modes from the associated
linearized hydrodynamic equations. This is quite an interesting
problem widely analyzed in the literature. As noted by the
different molecular dynamics experiments carried out for the USF
problem \cite{GT96,AL03,MD}, it becomes apparent the development
of inhomogeneities and formation of clusters as the flow
progresses. Consequently, the USF state is unstable for long
enough wavelength spatial perturbations. In order to understand
this phenomenon, several stability analysis have been undertaken
\cite{S92,B93,AN97,K00,K01}. Most of them are based on the
Navier-Stokes equations \cite{S92,B93,AN97} and, therefore, they
are limited to small velocity gradients, which for the USF problem
means small dissipation. Another alternative has been to solve the
Boltzmann equation by means of an expansion in a set of basis
functions \cite{K00,K01}. The coefficients of this expansion are
then determined by using also an expansion in powers of the
parameter $\epsilon \equiv \sqrt{1-\alpha^2}$, which is assumed to
be small. All these analytical results have shown that the USF becomes
unstable for certain kind of disturbances.
My approach is different from previous works since the
conditions for stability are obtained from a linear stability
analysis involving the transport coefficients of the {\em
perturbed} USF state instead of the usual Navier-Stokes
coefficients. Furthermore, the analysis is not restricted to the
low-dissipation limit since the reference state goes beyond this
range of values of $\alpha$. Two different perturbations to the
reference state have been considered here: (i) perturbations along
the velocity gradient ($y$ direction) only and (ii) perturbations along
the vorticity direction ($z$ direction) only. The results show that
the USF is linearly stable in the first case while it becomes
unstable in the second case. These results agree qualitatively with those
previously derived \cite{S92,AN97} in the context of the
Navier-Stokes description.
On the other hand, at a quantitative level, the comparison carried out here
shows significant differences between the Navier-Stokes description and
the present results as the collisions become more inelastic. In addition,
our results also confirm that the
instability is confined to long wavelengths (small wave numbers)
and so it can be avoided for small enough systems.

The plan of the paper is as follows. In Sec.\ \ref{sec2}, the
Boltzmann kinetic equation is introduced and a brief summary of
relevant results concerning the USF problem is given. In Sec.\
\ref{sec3}, the problem we are interested in is described and the
set of generalized transport coefficients characterizing the
transport around USF is defined. Explicit expressions for these
coefficients are provided in Sec.\ \ref{sec4} by using a kinetic
model of the Boltzmann equation. The details of the calculations
are displayed along several Appendices. Section \ref{sec5} is
devoted to the linear stability analysis around the steady USF
state and presents the form of the hydrodynamic modes. The paper
is closed in Sec.\ \ref{sec6} with a discussion of the results
obtained here.

\section{Boltzmann kinetic equation and uniform shear flow}
\label{sec2}

Let us consider a granular gas composed by smooth spheres of mass
$m$ and diameter $\sigma$. The inelasticity of collisions among
all pairs is accounted for by a {\em constant} coefficient of
restitution $0\leq \alpha \leq 1$ that only affects the
translational degrees of freedom of grains. In a kinetic theory
description all the relevant information on the state of the
system is given by the one-particle velocity distribution function
$f({\bf r}, {\bf v},t)$. At low density the inelastic Boltzmann
equation \cite{GS95,BDS97} gives the time evolution of $f({\bf r},
{\bf v},t)$. In the absence of an external force, it has the form
\begin{equation}
\label{2.1}
\left(\frac{\partial}{\partial t}+{\bf v}\cdot \nabla \right)f({\bf r}, {\bf v},t)=J[{\bf v}|f(t),f(t)],
\end{equation}
where the Boltzmann collision operator is
\begin{eqnarray}
\label{2.2}
J\left[{\bf v}_{1}|f,f\right]&=&\sigma^2 \int d{\bf v}_{2}
\int d\widehat{\boldsymbol {\sigma }}\,\Theta (\widehat{{\boldsymbol {\sigma }}}
\cdot {\bf g})(\widehat{\boldsymbol {\sigma}}\cdot {\bf g})\nonumber\\
& & \times
\left[ \alpha ^{-2} f({\bf r},{\bf v}_1')f({\bf r},{\bf v}_2',t)
-f({\bf r},{\bf v}_1,t)f({\bf r},{\bf v}_2,t)\right].
\end{eqnarray}
Here, $\widehat{\boldsymbol {\sigma}}$ is a unit vector along their line
of centers, $\Theta $ is
the Heaviside step function, and ${\bf g}={\bf v}_{1}-{\bf v}_{2}$ is the relative velocity.
The primes on the velocities denote the initial values $\{{\bf v}_{1}^{\prime },
{\bf v}_{2}^{\prime }\}$ that lead to $\{{\bf v}_{1},{\bf v}_{2}\}$
following a binary collision:
\begin{equation}
\label{2.3}
{\bf v}_{1}^{\prime}={\bf v}_{1}-\frac{1}{2}\left(1+\alpha^{-1}\right)
(\widehat{\boldsymbol {\sigma}}\cdot {\bf g})\widehat{\boldsymbol {\sigma}},
\quad {\bf v}_{2}^{\prime }={\bf v}_{2}+\frac{1}{2}\left(
1+\alpha^{-1}\right) (\widehat{\boldsymbol {\sigma}}\cdot
{\bf g})\widehat{\boldsymbol {\sigma}}
\end{equation}
The first five velocity moments of $f$ define the number density
\begin{equation}
\label{2.4}
n({\bf r}, t)=\int \; d{\bf v}f({\bf r}, {\bf v},t),
\end{equation}
the flow velocity
\begin{equation}
\label{2.5}
{\bf u}({\bf r}, t)=\frac{1}{n({\bf r}, t)}\int \; d{\bf v} {\bf v} f({\bf r},{\bf v},t),
\end{equation}
and the {\em granular} temperature
\begin{equation}
\label{2.6}
T({\bf r}, t)=\frac{m}{3 n({\bf r}, t)}\int \; d{\bf v} V^2({\bf r}, t) f({\bf r},{\bf v},t),
\end{equation}
where ${\bf V}({\bf r},t)\equiv {\bf v}-{\bf u}({\bf r}, t)$ is the peculiar velocity.
The macroscopic balance equations for density $n$, momentum $m{\bf u}$, and
energy $\frac{3}{2}nT$ follow directly from Eq.\ ({\ref{2.1}) by multiplying
with $1$, $m{\bf v}$, and $\frac{1}{2}mv^2$ and integrating over ${\bf v}$:
\begin{equation}
\label{2.7}
D_{t}n+n\nabla \cdot {\bf u}=0\;,
\end{equation}
\begin{equation}
\label{2.8}
D_{t}u_i+(mn)^{-1}\nabla_j P_{ij}=0\;,
\end{equation}
\begin{equation}
\label{2.9}
D_{t}T+\frac{2}{3n}\left(\nabla \cdot {\bf q}+P_{ij}\nabla_j u_i\right)
=-\zeta T\;,
\end{equation}
where $D_{t}=\partial _{t}+{\bf u}\cdot \nabla$. The microscopic
expressions for the pressure tensor ${\sf P}$, the heat flux ${\bf
q}$, and the cooling rate $\zeta$ are given, respectively, by
\begin{equation}
{\sf P}({\bf r}, t)=\int d{\bf v}\,m{\bf V}{\bf V}\,f({\bf r},{\bf v},t),
 \label{2.10}
\end{equation}
\begin{equation}
{\bf q}({\bf r}, t)=\int d{\bf v}\,\frac{1}{2}m V^{2}{\bf V}\,
f({\bf r},{\bf v},t),
\label{2.11}
\end{equation}
\begin{equation}
\label{2.12}
\zeta({\bf r}, t)=-\frac{1}{3n({\bf r},t)T({\bf r}, t)}\int\, d{\bf v} mV^2J[{\bf r},{\bf v}|f(t)].
\end{equation}

We assume that the gas is under uniform (or simple) shear flow (USF).
This idealized macroscopic
state is characterized by a constant density, a uniform temperature and a
simple shear with the local velocity field given by
\begin{equation}
\label{2.13} u_i=a_{ij}r_j, \quad a_{ij}=a\delta_{ix}\delta_{jy},
\end{equation}
where $a$ is the {\em constant} shear rate. This linear velocity
profile assumes no boundary layer near the walls and is generated
by the Lee-Edwards boundary conditions \cite{LE72}, which are
simply periodic boundary conditions in the local Lagrangian frame
moving with the flow velocity. For elastic gases, the temperature
grows in time due to viscous heating and so a steady state is not
possible unless an external (artificial) force is introduced
\cite{GS03}. However, for inelastic gases, the temperature changes
in time due to the competition between two (opposite) mechanisms:
on the one hand, viscous (shear) heating and, on the other hand,
energy dissipation in collisions. A steady state is achieved when
both mechanisms cancel each other and the fluid autonomously seeks
the temperature at which the above balance occurs. Under these
conditions, in the steady state the balance equation (\ref{2.9})
becomes
\begin{equation}
\label{2.14}
aP_{xy}=-\frac{3}{2}\zeta p,
\end{equation}
where $p=nT$ is the hydrostatic pressure. Note that for given values of
the shear rate $a$ and the coefficient of restitution $\alpha$, the
relation (\ref{2.14})
gives the temperature $T$ in the steady state as a {\em unique } function
of the density $n$.

The USF problem is perhaps the nonequilibrium state most widely
studied in the past few years both for granular and conventional
gases \cite{GS03,C90}. At a microscopic level, it becomes
spatially homogeneous when the velocities of the particles are
referred to the Lagrangian frame of reference co-moving with the
flow velocity ${\bf u}$ \cite{DSBR86}. Therefore, the one-particle
distribution function adopts the {\em uniform} form, $f({\bf
r},{\bf v})\rightarrow f({\bf V})$, and the Boltzmann equation
(\ref{2.1}) reads
\begin{equation}
\label{2.15}
-aV_y\frac{\partial}{\partial V_x} f({\bf V})=J\left[ {\bf V}|f,f\right] \;.
\end{equation}
This equation is invariant under the transformations
\begin{equation}
\label{2.15.1} V_z\rightarrow -V_z, \quad (V_x, V_y)\rightarrow
-(V_x, V_y),\quad (V_x, a)\rightarrow (-V_x, -a).
\end{equation}

The elements of the pressure tensor provide information on the
relevant transport properties of the USF problem. These elements
can be obtained by multiplying the Boltzmann equation (\ref{2.15})
by $m V_i V_j$ and integrating over ${\bf V}$. The result is
\begin{eqnarray}
\label{2.16}
a_{i\ell}P_{j\ell}+a_{j\ell}P_{i\ell}&=&m \int d{\bf V} V_iV_j
J[{\bf V}|f,f]\nonumber\\
&\equiv&\Lambda_{ij}.
\end{eqnarray}
The exact expression of the collision integral $\Lambda_{ij}$ is not known, even in
the elastic case. However, a good estimate can be expected by using Grad's approximation:
\begin{equation}
\label{2.17}
f({\bf V})\to f_{0}({\bf
V})\left[1+\frac{m}{2T}\left(\frac{P_{ij}}{p}-\delta_{ij}\right)V_iV_j\right],
\end{equation}
where $f_{0}({\bf V})$
\begin{equation}
\label{2.18}
f_{0}({\bf V})=n(m/2\pi T)^{3/2} \exp(-m V^2/2T)
\end{equation}
is the local equilibrium distribution function. When Eq.\ (\ref{2.17}) is substituted
into the definition of $\Lambda_{ij}$ and nonlinear terms in $P_{ij}/nT-\delta_{ij}$
are neglected, one gets \cite{G02}
\begin{equation}
\label{2.19} \Lambda_{ij}=-\nu\left[\beta
\left(P_{ij}-p\delta_{ij}\right)+\zeta^*P_{ij}\right],
\end{equation}
where
\begin{equation}
\label{2.20}
\nu(T)=\frac{16}{5}n\sigma^2\sqrt{\frac{\pi T}{m}},
\end{equation}
is an effective collision frequency,
\begin{equation}
\label{2.21}
\zeta^*=\frac{\zeta}{\nu}=\frac{5}{12}(1-\alpha^2),
\end{equation}
is the dimensionless cooling rate evaluated in the local equilibrium approximation and
\begin{equation}
\label{2.22}
\beta=\frac{1+\alpha}{2}\left(1-\frac{1-\alpha}{3}\right).
\end{equation}
The set of coupled equations for $P_{ij}$ can be now easily solved when one
takes into account the approach (\ref{2.19}).
The expressions for the reduced elements $P_{ij}^*=P_{ij}/p$ are
\begin{equation}
\label{2.23}
P_{xx}^*=3-2P_{yy}^*,\quad P_{yy}^*=P_{zz}^*=\frac{\beta}{\beta+\zeta^*},
\quad P_{xy}^*=-\frac{\beta}{(\beta+\zeta^*)^2}a^*,
\end{equation}
where the (reduced) shear rate $a^*=a/\nu$ is given by
\begin{equation}
\label{2.24}
a^*=\sqrt{\frac{3}{2}\frac{\zeta^*}{\beta}}\left(\beta+\zeta^*\right).
\end{equation}
The expression (\ref{2.24}) clearly indicates the intrinsic connection between the
(reduced) velocity gradient and dissipation in the system. In fact, in the elastic limit
($\alpha=1$, which implies $a^*=0$), the equilibrium results of the ordinary gas are
recovered, i.e., $P_{ij}^*=\delta_{ij}$. This means that $\alpha$ (or $a^*$) can be considered
as the relevant nonequilibrium parameter of the problem.
The analytical results given by Eqs.\ (\ref{2.23}) and (\ref{2.24}) agree quite
well \cite{SGD04,MG02} with Monte Carlo simulations of the Boltzmann equation
\cite{MG02,AS05}, even for strong dissipation.

\section{Small perturbations from the uniform shear flow: Transport coefficients}
\label{sec3}

In general, the USF state can be disturbed by small spatial perturbations. The response of
the system to these perturbations gives rise to additional contributions to the momentum
and heat fluxes,
which can be characterized by generalized transport coefficients.
This section is devoted to the study of such small perturbations.

In order to analyze this problem we have to start from the Boltzmann equation
with a general time and space dependence. Let ${\bf u}_0={\sf a}\cdot {\bf r}$ be the
flow velocity of the {\em undisturbed} USF state. Here, the only nonzero element
of the tensor ${\sf a}$ is $a_{ij}=a\delta_{ix}\delta_{jy}$.
In the {\em disturbed} state, however
the true velocity ${\bf u}$ is in general different from ${\bf u}_0$ since
${\bf u}={\bf u}_0+\delta {\bf u}$, $\delta {\bf u}$ being a small perturbation to ${\bf u}_0$.
As a consequence, the true peculiar velocity
is now  ${\bf c}\equiv {\bf v}-{\bf u}={\bf V}-\delta{\bf u}$, where ${\bf V}={\bf v}-{\bf u}_0$.
In the Lagrangian frame moving with ${\bf u}_0$, the Boltzmann equation can be written as
\begin{equation}
\label{3.1}
\frac{\partial}{\partial
t}f-aV_y\frac{\partial}{\partial V_x}f+\left({\bf V}+{\bf
u}_0\right) \cdot \nabla f=J[{\bf V}|f,f],
\end{equation}
where here the derivative $\nabla f$ is taken at constant ${\bf V}$. The corresponding macroscopic balance
equations associated with this disturbed USF state follows from the general equations (\ref{2.7})--(\ref{2.9})
when one takes into account that ${\bf u}={\bf u}_0+\delta {\bf u}$. The result is
\begin{equation}
\label{3.2}
\partial_tn+{\bf u}_0\cdot \nabla n=-\nabla \cdot (n\delta {\bf u}),
\end{equation}
\begin{equation}
\label{3.3}
\partial_t\delta {\bf u}+{\sf a}\cdot \delta {\bf u}+({\bf u}_0+\delta {\bf u})\cdot \nabla \delta {\bf u}=-
(mn)^{-1}\nabla \cdot {\sf P},
\end{equation}
\begin{equation}
\label{3.4}
\frac{3}{2}n\partial_tT+\frac{3}{2}n({\bf u}_0+\delta {\bf u})\cdot \nabla T+aP_{xy}+\nabla
\cdot {\bf q}+{\sf P}:\nabla \delta {\bf u}=-\frac{3}{2}p\zeta,
\end{equation}
where the pressure tensor ${\sf P}$, the heat flux ${\bf q}$ and
the cooling rate $\zeta$ are defined by Eqs.\
(\ref{2.10})--(\ref{2.12}), respectively, with the replacement
${\bf V}\rightarrow {\bf c}$.

We assume now that the deviations from the USF state are small,
which means that the spatial gradients of the hydrodynamic fields
\begin{equation}
\label{3.5}
A({\bf r},t)\equiv \{n({\bf r},t), T({\bf r}, t), \delta {\bf u}({\bf r},t)\}
\end{equation}
are small. Under these conditions, a solution to the Boltzmann
equation (\ref{3.1}) can be obtained by means of a generalization
of the conventional Chapman-Enskog method \cite{CC70} where the
velocity distribution function is expanded about a {\em local}
shear flow reference state in terms of the small spatial gradients
of the hydrodynamic fields relative to those of USF. This type of
Chapman-Enskog-like expansion has been considered in the case of
elastic gases to get the set of shear-rate dependent transport
coefficients \cite{GS03,LD97} in a thermostatted shear flow
problem and it has also been recently considered \cite{L05} in the
context of inelastic gases.

To construct the Chapman-Enskog expansion let us look for a {\em
normal} solution of the form
\begin{equation}
\label{3.6}
f({\bf r}, {\bf V},t)\equiv f(A({\bf r}, t), {\bf V}).
\end{equation}
This special solution expresses the fact that the space dependence of the reference
shear flow is completely absorbed in the relative velocity ${\bf V}$ and
all other space and time dependence occurs entirely through a functional dependence
on the fields $A({\bf r}, t)$. The functional dependence can be made local by an
expansion of the distribution function in powers of the hydrodynamic gradients:
\begin{equation}
\label{3.7}
f({\bf r}, {\bf V},t)
=f^{(0)}(A({\bf r}, t), {\bf V})+ f^{(1)}(A({\bf r}, t), {\bf V})+\cdots,
\end{equation}
where the reference zeroth-order distribution function corresponds
to the USF distribution function but taking into account the local
dependence of the density and temperature and the change ${\bf
V}\rightarrow {\bf V}-\delta{\bf u}({\bf r}, t)$ [see Eqs.\ (\ref{b3}) and
(\ref{b4}) for the explicit form of $f^{(0)}$ in the steady state
given by a kinetic
model of the Boltzmann equation]. The successive approximations
$f^{(k)}$ are of order $k$ in the gradients of $n$, $T$, and
$\delta {\bf u}$ but retain all the orders in the shear rate $a$.
This is the main feature of this expansion. In this paper, only
the first order approximation will be considered. More details on this
Chapman-Enskog-like type of expansion can be found in Ref.\ \cite{L05}.

The expansion (\ref{3.7}) yields the corresponding expansion for the fluxes and the cooling rate when
one substitutes (\ref{3.7}) into their definitions (\ref{2.10})--(\ref{2.12}):
\begin{equation}
\label{3.8}
{\sf P}={\sf P}^{(0)}+{\sf P}^{(1)}+\cdots, \quad {\bf
q}={\bf q}^{(0)}+{\bf q}^{(1)}+\cdots, \quad
\zeta=\zeta^{(0)}+\zeta^{(1)}+\cdots.
\end{equation}
Finally, as in the usual Chapman-Enskog method, the time derivative is also expanded as
\begin{equation}
\label{3.9}
\partial_t=\partial_t^{(0)}+\partial_t^{(1)}+\partial_t^{(2)}+\cdots,
\end{equation}
where the action of each operator $\partial_t^{(k)}$ is obtained from the hydrodynamic equations
(\ref{3.2})--(\ref{3.4}). These results provide the basis for generating the Chapman-Enskog
solution to the inelastic Boltzmann equation (\ref{3.1}).

\subsection{Zeroth-order approximation}

Substituting the expansions (\ref{3.7}) and (\ref{3.9}) into Eq.\ (\ref{3.1}), the kinetic equation
for $f^{(0)}$ is given by
\begin{equation}
\label{n1}
\partial_t^{(0)}f^{(0)}-aV_y\frac{\partial}{\partial V_x}f^{(0)}=J[{\bf V}|f^{(0)},f^{(0}].
\end{equation}
To lowest order in the expansion the conservation laws give
\begin{equation}
\label{3.10}
\partial_t^{(0)}n=0,\quad \partial_t^{(0)}T=-\frac{2}{3n}a P_{xy}^{(0)}-T\zeta^{(0)},
\end{equation}
\begin{equation}
\label{3.11}
\partial_t^{(0)}\delta u_i+a_{ij} \delta u_j=0.
\end{equation}
As said before, for given values of $a$ and $\alpha$, the steady
state condition (\ref{2.14}) establishes a mapping between the
density and temperature so that every density corresponds to one
and only one temperature. Since the density $n({\bf r}, t)$ and
temperature $T({\bf r}, t)$ are specified separately in the {\em
local} USF state, the viscous heating only partially compensates
for the collisional cooling and so, $\partial_t^{(0)} T \neq 0$.
Consequently, the zeroth-order distribution $f^{(0)}$ depends on
time through its dependence on the temperature. Because of the
steady state condition (\ref{2.14}) does not apply in general
locally, the reduced shear rate $a^*=a/\nu(n,T)$ depends on space
and time so that, $a^*$ and $\alpha$ must be considered as
independent parameters for general infinitesimal perturbations
around the USF state. Since $f^{(0)}$ is a normal solution, then
\begin{eqnarray}
\label{n3}
\partial_t^{(0)}f^{(0)}&=&\frac{\partial f^{(0)}}{\partial
n}\partial_t^{(0)} n+\frac{\partial f^{(0)}}{\partial
T}\partial_t^{(0)} T+\frac{\partial f^{(0)}}{\partial \delta
u_i}\partial_t^{(0)} \delta u_i\nonumber\\
&=&-\left(\frac{2}{3n}a
P_{xy}^{(0)}+T\zeta^{(0)}\right)\frac{\partial}{\partial T}f^{(0)}-a_{ij}\delta u_j
\frac{\partial}{\partial \delta u_i}f^{(0)}\nonumber\\
&=&-\left(\frac{2}{3n}a
P_{xy}^{(0)}+T\zeta^{(0)}\right)\frac{\partial}{\partial T}f^{(0)}+a_{ij}\delta u_j
\frac{\partial}{\partial c_i}f^{(0)},
\end{eqnarray}
where in the last step we have taken into account that $f^{^(0)}$ depends on $\delta
{\bf u}$ only through the peculiar velocity ${\bf c}$.
Substituting (\ref{n3}) into (\ref{n1}) yields the following kinetic equation for $f^{(0)}$:
\begin{equation}
\label{n4}
-\left(\frac{2}{3n}a
P_{xy}^{(0)}+T\zeta^{(0)}\right)\frac{\partial}{\partial T}f^{(0)}
-ac_y\frac{\partial}{\partial c_x}f^{(0)}=J[{\bf V}|f^{(0)},f^{(0}].
\end{equation}
The zeroth-order solution leads to ${\bf q}^{(0)}={\bf 0}$. On the other hand,
to solve Eq.\ (\ref{n4}) one needs to know the temperature dependence of the zeroth momentum
flux $P_{xy}^{(0)}$. A closed set of equations for ${\sf P}^{(0)}$ is obtained when one considers
Grad's approximation (\ref{2.17}):
\begin{equation}
\label{n5}
-\left(\frac{2}{3n}a
P_{xy}^{(0)}+T\zeta^{(0)}\right)\frac{\partial}{\partial T} P_{ij}^{(0)}+
a_{i\ell}P_{j\ell}^{(0)}+a_{j\ell}P_{i\ell}^{(0)}=
-\nu\left[\beta
\left(P_{ij}^{(0)}-p\delta_{ij}\right)+\zeta^*P_{ij}^{(0)}\right],
\end{equation}
where
\begin{equation}
\label{3.13}
\zeta^*=\frac{\zeta^{(0)}}{\nu}=\frac{5}{12}(1-\alpha^2).
\end{equation}
The steady state solution of Eq.\ (\ref{n5}) is given by Eqs.\ (\ref{2.23})
and (\ref{2.24}). However, in general the equations (\ref{n5}) must be solved numerically to get
the dependence of the zeroth-order pressure tensor $P_{ij}^{(0)}(T)$ on temperature.
A detailed study on the unsteady hydrodynamic solution of Eqs.\ (\ref{n5})
has been carried out in Ref.\ \cite{SGD04}.
In what follows, $P_{ij}^{(0)}(T)$ will be considered as a known function of $T$.

\subsection{First-order approximation}

The analysis to first order in the gradients is worked out in
Appendix \ref{appA}. Only the final results are presented in this
Section. The distribution function $f^{(1)}$ is of the form
\begin{equation}
\label{3.14}
f^{(1)}={\bf X}_{n}\cdot \nabla n+ {\bf X}_{T}\cdot \nabla T+{\sf X}_{u}:\nabla \delta {\bf u},
\end{equation}
where the vectors ${\bf X}_{n}$ and ${\bf X}_{T}$ and the tensor ${\sf X}_{u}$ are functions of
the true peculiar velocity ${\bf c}$. They are the
solutions of the following linear integral equations:
\begin{equation}
\label{3.15}
-\left[\left(\frac{2}{3n}a
P_{xy}^{(0)}+T\zeta^{(0)}\right)\partial_T+ a
c_y\frac{\partial}{\partial c_x}-{\cal
L}\right]X_{n,i}+\frac{T}{n}\left[\frac{2a}{3p}(1-n\partial_n)
P_{xy}^{(0)}-\zeta^{(0)}\right]X_{T,i}=Y_{n,i},
\end{equation}
\begin{equation}
\label{3.16} -\left[\left(\frac{2}{3n}a
P_{xy}^{(0)}+T\zeta^{(0)}\right)\partial_T+
\frac{2a}{3p}T(\partial_T P_{xy}^{(0)})+\frac{3}{2}\zeta^{(0)}+a
c_y\frac{\partial}{\partial c_x}-{\cal L}\right]X_{T,i}=Y_{T,i},
\end{equation}
\begin{equation}
\label{3.17}
-\left[\left(\frac{2}{3n}a
P_{xy}^{(0)}+T\zeta^{(0)}\right)\partial_T+a
c_y\frac{\partial}{\partial c_x}-{\cal L}\right]X_{u,k\ell}-
a\delta_{ky}X_{u,x\ell}-\zeta_{u,k\ell}T\partial_T
f^{(0)}=Y_{u,k\ell},
\end{equation}
where ${\bf Y}_n({\bf c})$, ${\bf Y}_T({\bf c})$, and ${\sf
Y}_u({\bf c})$ are defined by Eqs.\ (\ref{a9})--(\ref{a11}),
respectively, and $\zeta_{u,k\ell}$ is defined by Eq.\
(\ref{a12.2}). While the $Y$ functions are given in terms of the
reference state distribution $f^{(0)}$, $\zeta_{u,k\ell}$ is a
functional of the unknown $X_{u,k\ell}$. In addition, ${\cal L}$
is the linearized Boltzmann collision operator around the
reference state
\begin{equation}
\label{3.17bis}
{\cal L}X\equiv
-\left(J[f^{(0)},X]+J[X,f^{(0)}]\right).
\end{equation}
A good estimate of $\zeta_{u,k\ell}$ can be obtained by expanding
$X_{u,k\ell}$ in a complete set of polynomials (for instance,
Sonine polynomials) and then truncating the series after the first
few terms. In practice, the leading term in these expansions
provides a very accurate result over a wide range of dissipation.
This contribution has been obtained in Appendix \ref{appC} and is
given by Eq.\ (\ref{c10}).

With the distribution function $f^{(1)}$ determined by (\ref{3.14}), the first-order
corrections to the fluxes are
\begin{equation}
\label{3.18}
P_{ij}^{(1)}=-\eta_{ijk\ell} \frac{\partial \delta u_k}
{\partial r_{\ell}},
\end{equation}
\begin{equation}
\label{3.19}
q_i^{(1)}=-\kappa_{ij}\frac{\partial T}{\partial r_j}-
\mu_{ij}\frac{\partial n}{\partial r_j},
\end{equation}
where
\begin{equation}
\label{3.20}
\eta_{ijk\ell}=-\int\; d{\bf c}\, mc_ic_j X_{u,k\ell}({\bf c}),
\end{equation}
\begin{equation}
\label{3.21}
\kappa_{ij}=-\int\; d{\bf c}\, \frac{m}{2}c^2c_i X_{T,j}({\bf c}),
\end{equation}
\begin{equation}
\label{3.22}
\mu_{ij}=-\int\; d{\bf c}\, \frac{m}{2}c^2c_i X_{n,j}({\bf c}).
\end{equation}
Upon writing Eqs.\ (\ref{3.18})--(\ref{3.22}) use has been made of
the symmetry properties of $X_{n,i}$ $X_{T,i}$ and $X_{u,ij}$. In
general, the set of {\em generalized} transport coefficients
$\eta_{ijk\ell}$, $\kappa_{ij}$, and $\mu_{ij}$ are nonlinear
functions of the coefficient of restitution $\alpha$ and the
reduced shear rate $a^*$. The anisotropy induced in the system by
the shear flow gives rise to new transport coefficients,
reflecting broken symmetry. The momentum flux is expressed in
terms of a viscosity tensor $\eta_{ijk\ell}(\alpha)$ of rank 4
which is symmetric and traceless in $ij$ due to the properties of
the pressure tensor $P_{ij}^{(1)}$. The heat flux is expressed in
terms of a thermal conductivity tensor $\kappa_{ij}(\alpha)$ and a
new tensor $\mu_{ij}(\alpha)$.

\subsection{Steady state conditions}

As shown in the above subsections, the evaluation of the complete
nonlinear dependence of the generalized transport coefficients on
the shear rate and dissipation requires the analysis of the
hydrodynamic behavior of the {\em unsteady} reference state. This
involves the corresponding numerical integrations of the
differential equations obeying the velocity moments of the
zeroth-order solution. This is a quite intricate and long problem.
However, given that here we are mainly interested in performing a
linear stability analysis of the hydrodynamic equations with
respect to the steady state, we want to evaluate the transport
coefficients in this special case.
As a consequence, $\partial_t^{(0)}T=0$ and so the condition
\begin{equation}
\label{3.12}
a^*P_{xy}^{*}=-\frac{3}{2}\zeta^{*}.
\end{equation}
applies. In Eq.\ (\ref{3.12}), it is understood that $a^*$ and $P_{xy}^*=P_{xy}^{(0)}/p$
are evaluated in the steady state, namely, they are given by Eqs.\ (\ref{2.23}) and
(\ref{2.24}), respectively.
A consequence of Eq.\ (\ref{3.12}) is that the first term on the left hand
side of the integral equations (\ref{3.15})--(\ref{3.17})
vanishes. In addition, the dependence of the pressure tensor
$P_{ij}^{(0)}$ on density and
temperature occurs explicitly through $p=nT$ and through its dependence on
$a^*$. In this case, the derivatives $\partial_n P_{ij}^{(0)}$ and $\partial_T P_{ij}^{(0)}$ can be
written more explicitly as
\begin{equation}
\label{n8}
n\partial_n P_{ij}^{(0)}=n\partial_n p P_{ij}^*(a^*)=p\left(1-a^*\frac{\partial}
{\partial a^*}\right)P_{ij}^*(a^*),
\end{equation}
\begin{equation}
\label{n9}
T\partial_T P_{ij}^{(0)}=T\partial_T p P_{ij}^*(a^*)=p\left(1-\frac{1}{2}a^*\frac{\partial}
{\partial a^*}\right)P_{ij}^*(a^*).
\end{equation}
The dependence of $P_{ij}^*$ on $a^*$ near the steady state is
determined in the Appendix \ref{appD} so that, all the terms
appearing in the integral equations are explicitly known in the
steady state. Under the above conditions, Eqs.\
(\ref{3.15})--(\ref{3.17}) become
\begin{equation}
\label{3.15bis} \left(-a c_y\frac{\partial}{\partial c_x}+{\cal
L}\right)X_{n,i}+\frac{2a}{3}\frac{T}{n}(P_{xy}^*+a^*\partial_a^*
P_{xy}^{*})X_{T,i}=Y_{n,i},
\end{equation}
\begin{equation}
\label{3.16bis}
 \left(-a c_y\frac{\partial}{\partial c_x}-\frac{1}{3}a\left(
 P_{xy}^*-a^*\partial_{a^*}P_{xy}^*\right)+{\cal L}\right)X_{T,i}=Y_{T,i},
\end{equation}
\begin{equation}
\label{3.17bis}
\left(-a c_y\frac{\partial}{\partial c_x}+{\cal
L}\right)X_{u,k\ell}-
a\delta_{ky}X_{u,x\ell}-\zeta_{u,k\ell}T\partial_T
f^{(0)}=Y_{u,k\ell},
\end{equation}
where it is understood again that in Eqs.\
(\ref{3.15bis})--(\ref{3.17bis}) all the quantities are evaluated
in the steady state. Henceforth, I will restrict my calculations
to this particular case.

Given that in the steady state the coefficient of restitution and
the reduced shear rate are coupled, the usual Navier-Stokes
transport coefficients for ordinary gases are recovered for
elastic collisions ($a^*=0$). Thus, when $\alpha \to 1$ the coefficients become
\begin{equation}
\label{3.23} \eta_{ijk\ell}\rightarrow
\eta_0\left(\delta_{ik}\delta_{j\ell}+\delta_{jk}\delta_{i\ell}-
\frac{2}{3}\delta_{ij}\delta_{k\ell}\right),\quad
\kappa_{ij}\rightarrow \kappa_0 \delta_{ij}, \quad
\mu_{ij}\rightarrow 0,
\end{equation}
where $\eta_0=p/\nu$ and $\kappa_0=15 \eta_0/4m$ are the shear
viscosity and thermal conductivity coefficients given by the
(elastic) Boltzmann equation.

\section{Results from a simple kinetic model}
\label{sec4}

The explicit form of the generalized transport coefficients $\mu_{ij}$, $\kappa_{ij}$ and
$\eta_{ijk\ell}$
requires to solve the integral equations
(\ref{3.15bis})--((\ref{3.17bis}). Apart from the mathematical
difficulties embodied in the Boltzmann collision operator ${\cal
L}$, the fourth-degree velocity moments of the distribution
$f^{(0)}$ are also needed to determine $\mu_{ij}$ and $\kappa_{ij}$
and they are not provided in principle by the Grad
approximation. Nevertheless, an accurate estimate of these moments
from the Boltzmann equation is a formidable task since it would
require at least to include the fourth-degree moments in Grad's
solution. In this case, to overcome such difficulties it is useful
to consider a model kinetic equation of the Boltzmann equation. As
for elastic collisions, the idea is to replace the true Boltzmann
collision operator with a simpler, more tractable operator that
retains the most relevant physical properties of the Boltzmann
operator. Here, I consider a kinetic model \cite{BDS99} based on
the well-known Bhatnagar-Gross-Krook (BGK) \cite{GS03} for
ordinary gases where the operator $J[f,f]$ is \cite{note}
\begin{equation}
\label{4.1}
J[f,f]\to -\beta \nu (f-f_0)+\frac{\zeta}{2}\frac{\partial}
{\partial {\bf c}}\cdot \left({\bf c}f\right).
\end{equation}
Here, $\nu$ and $\beta$ are given by Eqs.\ (\ref{2.20}) and
(\ref{2.22}), respectively, $f_0$ is the local equilibrium
distribution (\ref{2.18}) and $\zeta$ is the cooling rate defined
by Eq.\ (\ref{2.12}). As said before, an estimate of $\zeta$ to
first order in the gradients has been derived in Appendix
\ref{appC}. In general, the quantity $\beta$ can be considered as
an adjustable parameter to optimize the agreement with the
Boltzmann equation. In this paper, $\beta$ has been chosen to
reproduce the true Navier-Stokes shear viscosity coefficient of an
inelastic gas of hard spheres \cite{single}. A slightly different
choice for $\beta$, namely $\beta=(1+\alpha)/2$, is considered in
Ref.\ \cite{AS05}.
\begin{figure}
\includegraphics[width=0.5 \columnwidth,angle=0]{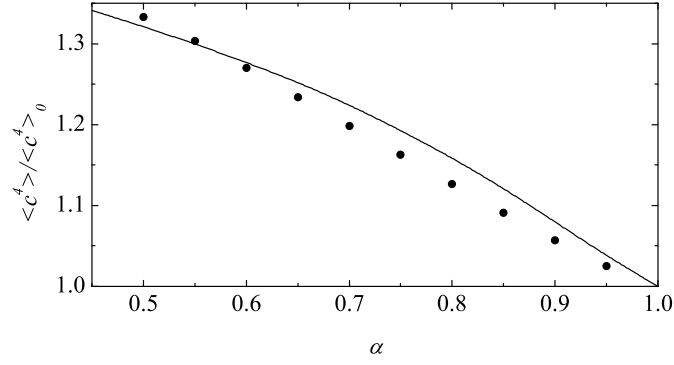}
\caption{Fourth-degree velocity moment $\langle c^4 \rangle$
relative to its local equilibrium value as a function of the
coefficient of restitution. Solid line is the prediction of the
kinetic model while the symbols are simulation results
\cite{AS05}.
\label{fig6}}
\end{figure}

By taking moments with respect to $1$, ${\bf c}$ and $c^2$, the
model kinetic equation (\ref{4.1}) yields the same form of the
macroscopic balance equations for mass, momentum, and energy,
Eqs.\ (\ref{2.7})--(\ref{2.9}), as those given from the Boltzmann
equation. When $\alpha=1$, then $\beta=1$, $\zeta=0$ and so the
kinetic model (\ref{4.1}) reduces to the BGK equation whose
utility to address complex states not accessible via the Boltzmann
equation is well-established for elastic gases \cite{GS03}. In the
case of granular gases, it is easy to show that the kinetic model
leads to the same results for the pressure tensor in the USF
problem as those given from Grad's solution to the Boltzmann
equation, Eqs.\ (\ref{2.23})--(\ref{2.24}). This result, along
with those of Refs.\ \cite{MGSB99} and \cite{TTMGSD01}, confirms
the reliability of the kinetic model for granular media as well. A
summary of the USF results derived from the kinetic model is
provided in Appendix \ref{appB}. In particular, beyond rheological
properties, recent computer simulations \cite{AS05} have confirmed
the accuracy of the kinetic model to capture the dependence of the
fourth-degree velocity moments (whose expressions are needed to
get the coefficients $\mu_{ij}$ and $\kappa_{ij}$ on dissipation
in the USF state. To illustrate it, in Fig.\ \ref{fig6} we plot
the fourth-degree moment
\begin{equation}
\label{b7}
\langle c^4 \rangle =\int\; d{\bf c} \;c^4 f({\bf c})
\end{equation}
relative to its local equilibrium value $\langle c^4
\rangle_{0}=15 nT^2/m^2$. The symbols refer to the numerical
results obtained from the DSMC method \cite{AS05}. It is quite
apparent that the analytical results agree well with simulation
data (the discrepancies between both results are smaller than
3\%), showing again that the reliability of the kinetic model goes
beyond the quasielastic limit.
\begin{figure}
\includegraphics[width=0.5 \columnwidth,angle=0]{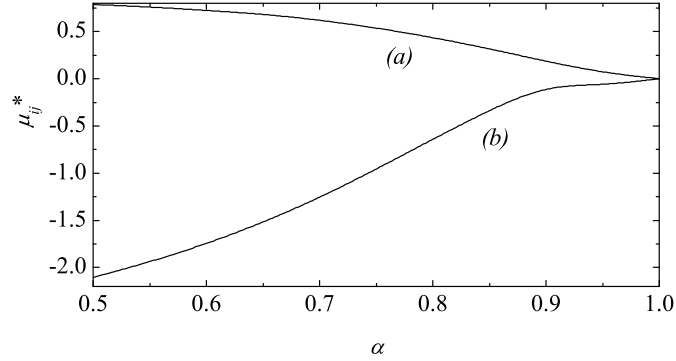}
\caption{Plot of the reduced coefficients $(a)$ $\mu_{yy}^*$ and
$(b)$ $\mu_{xy}^*$ as a function of the coefficient of restitution
$\alpha$. \label{fig1}}
\end{figure}
\begin{figure}
\includegraphics[width=0.5 \columnwidth,angle=0]{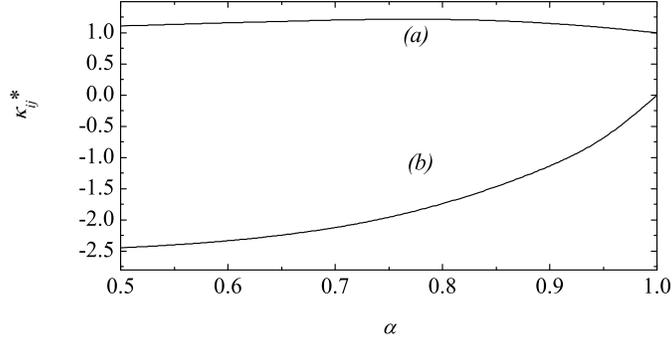}
\caption{Plot of the reduced coefficients $(a)$ $\kappa_{yy}^*$
and $(b)$ $\kappa_{xy}^*$ as a function of the coefficient of
restitution $\alpha$. \label{fig2}}
\end{figure}

Let us consider the perturbed USF problem in the context of the
kinetic model. By using the model (\ref{4.1}), the integral
equations (\ref{3.15bis})--(\ref{3.17bis}) still apply with the
only replacement
\begin{equation}
\label{4.2.1}
{\cal L}X\to \nu \beta X-\frac{\zeta^{(0)}}{2}
\frac{\partial}{\partial {\bf c}}\cdot \left({\bf c}X\right),
\end{equation}
in the case of $X_{n,i}$ and $X_{T,i}$ and
\begin{equation}
\label{4.2.2}
{\cal L}X_{ij}\to \nu \beta X_{ij}-\frac{\zeta^{(0)}}{2}
\frac{\partial}{\partial {\bf c}}\cdot \left({\bf c}X_{ij}\right)
-\frac{\zeta_{u,ij}}{2}\frac{\partial}{\partial {\bf c}}\cdot
\left({\bf c}f^{(0)}\right),
\end{equation}
in the case of $X_{u,ij}$. In the above equations, $\zeta^{(0)}$
is the zeroth-order approximation to $\zeta$ which is given by
Eq.\ (\ref{3.13}). With the changes (\ref{4.2.1}) and
(\ref{4.2.2}) all the generalized transport coefficients can be
easily evaluated from Eqs.\ (\ref{3.15bis})--(\ref{3.17bis}).
Details of these calculations are also given in Appendix
\ref{appC}; a more complete listing can be obtained on request
from the author.

The dependence of the generalized transport coefficients on the
coefficient of restitution $\alpha$ is illustrated in Figs.\ \ref{fig1},
\ref{fig2} and \ref{fig3} for the (reduced)
coefficients $\mu_{ij}^*$, $\kappa_{xy}^*$,
$\kappa_{yy}^*$, $\eta_{xxyy}^*$,  $\eta_{yyyy}^*$,
$\eta_{zzyy}^*$, and  $\eta_{xyyy}^*$. Here,
$\mu_{ij}^*=n\mu_{ij}/T\kappa_0$, $\kappa_{ij}^*=\kappa_{ij}/\kappa_0$
and
$\eta_{ijk\ell}^*=\eta_{ijk\ell}/\eta_0$, where $\eta_0=p/\nu$ and
$\kappa_0=5\eta_0/2m$ are the elastic values of the shear
viscosity and thermal conductivity coefficients given by the BGK
kinetic model. In general, we observe that the influence of
dissipation on the transport coefficients is quite significant.

\begin{figure}
\includegraphics[width=0.5 \columnwidth,angle=0]{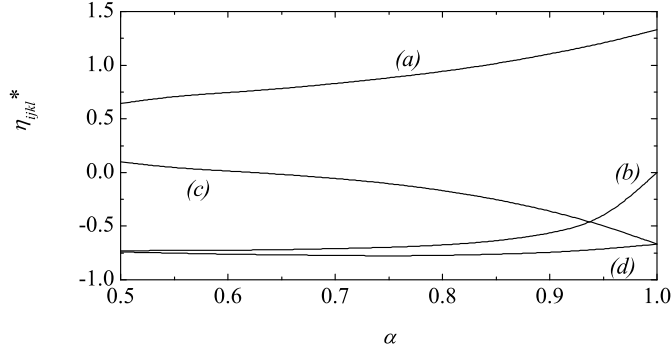}
\caption{Plot of the reduced coefficients $(a)$ $\eta_{yyyy}^*$,
$(b)$ $\eta_{xyyy}^*$, $(c)$ $\eta_{zzyy}^*$, and $(d)$
$\eta_{xxyy}^*$ as a function of the coefficient of restitution
$\alpha$. \label{fig3}}
\end{figure}

With all the transport coefficients known, the new constitutive
equations (\ref{3.18}) and (\ref{3.19}) are completed and the
corresponding set of closed hydrodynamic equations
(\ref{3.2})--(\ref{3.4}) can be derived. They are given by
\begin{equation}
\label{4.3}
\partial_tn+{\bf u}_0\cdot \nabla n+\nabla \cdot (n\delta {\bf u})
=0,
\end{equation}
\begin{equation}
\label{4.4}
\partial_t\delta u_i+a_{ij}\delta u_j+\left({\bf u}_0+\delta {\bf u}\right)\cdot \nabla \delta u_i+
\frac{1}{mn}\frac{\partial}{\partial r_j}\left(P_{ij}^{(0)}-\eta_{ijk\ell}
\frac{\partial \delta u_k}{\partial r_\ell}\right)=0,
\end{equation}
\begin{eqnarray}
\label{4.5}
& &\frac{3}{2}n\partial_t T+\frac{3}{2}n({\bf u}_0+\delta {\bf u})\cdot \nabla T-
a\eta_{xyij}\frac{\partial \delta u_i}{\partial r_j}\nonumber\\
&-&\frac{\partial}{\partial r_i}\left(\mu_{ij}\frac{\partial
n}{\partial r_j}+ \kappa_{ij}\frac{\partial T}{\partial
r_j}\right) +\left(P_{ij}^{(0)}- \eta_{ijk\ell}\frac{\partial
\delta u_k}{\partial r_\ell}\right)\frac{\partial \delta
u_i}{\partial r_j} +a P_{xy}^{(0)}\nonumber\\
&=&-\frac{3}{2}nT\zeta -\frac{3}{2}nT\zeta_{u,ij}\frac{\partial
\delta u_i}{\partial r_j}
\end{eqnarray}
Note also that consistency would require to consider the term
$aP_{xy}^{(2)}$ which is of second order in gradients and so, it
should be retained. Given that this would require to determine the
second order contributions to the fluxes, this term will be
neglected in our study. An important feature of our linearized
hydrodynamic equations is that they are not restricted to small
values of the (reduced) shear rate or, equivalently, to small
inelasticity. This allows us to go beyond the usual Navier-Stokes
hydrodynamics. The hydrodynamic equations (\ref{4.3})--(\ref{4.5})
are the starting point of the linear stability analysis of the USF
of the next Section.

\section{Linear stability analysis of the steady shear flow state}
\label{sec5}

As said in the Introduction, computer simulations \cite{MD} have
clearly shown that the USF state is unstable with respect to long
enough wavelength perturbations. These results have been also
confirmed by different analytical results \cite{S92,B93,AN97,K00},
most of them based on the Navier-Stokes description that applies
to first order in the shear rate. However, given that USF is
inherently non-Newtonian \cite{SGD04}, the full nonlinear
dependence of the transport coefficients on the shear rate is
required to perform a consistent linear stability analysis of the
nonlinear hydrodynamic equations (\ref{4.3})--(\ref{4.5}) with
respect to the USF state for small initial excitations. This
analysis allows one to determine the hydrodynamic modes for states
near USF as well the conditions for instabilities at long
wavelengths. A growth of these modes signals the onset of
instability, which is ultimately controlled by the dominance of
nonlinear terms. Note also that while all the works have been
mainly devoted to dense systems, much less attention has been paid
to dilute gases.

Let us assume that the deviations $\delta x_\mu ({\bf r},t)=x_\mu({\bf r},t)-x_{0\mu}({\bf r})$
are small, where $\delta x_\mu ({\bf r},t)$ denotes the deviation of $\{n, {\bf u}, T\}$ from their
values in the USF state $\{n_0, {\bf u}_0, T_0\}$. The quantities in the USF verify
\begin{equation}
\label{5.1} \nabla n_0=\nabla T_0=0,\quad {\bf u}_0={\sf a}\cdot
{\bf r},\quad \partial_t T_0=0.
\end{equation}
Now, let us linearize Eqs.\ (\ref{4.3})--(\ref{4.5}) with respect
to
\begin{equation}
\label{5.2}
\{\delta x_\mu ({\bf r},t)\}\equiv \left\{\delta n ({\bf r},t),
\delta T ({\bf r},t), \delta {\bf u} ({\bf r},t)\right\}.
\end{equation}
The resulting set of five linearized hydrodynamic equations
follows from Eqs.\ (\ref{4.3})--(\ref{4.5}):
\begin{equation}
\label{5.3}
\partial_t\delta n+ay\frac{\partial}{\partial x}\delta n+
n_0\cdot \delta {\bf u}=0,
\end{equation}
\begin{eqnarray}
\label{5.4} & & \frac{3}{2}n_0\partial_t \delta
T+ay\frac{\partial}{\partial x}\delta T+a\delta_{ix}\delta
u_y+a\left[(\partial_{n}P_{xy}^{(0)})\delta
n+(\partial_{T}P_{xy}^{(0)})\delta T\right]\nonumber\\
& & + \left( P_{k\ell}^{(0)}-a\eta_{xyk\ell}\right)\frac{\partial
\delta u_k}{\partial r_\ell}-\mu_{ij}\frac{\partial^2 \delta
n}{\partial r_{i}\partial r_{j}} -\kappa_{ij}\frac{\partial^2
\delta
T}{\partial r_i\partial r_j}\nonumber\\
&=&-\frac{3}{2}\zeta_0n_0T_0\left(2\frac{\delta
n}{n_0}+\frac{3}{2}\frac{\delta T}{T_0}\right)
-\frac{3}{2}n_0T_0\zeta_{u,k\ell}\frac{\partial \delta
u_k}{\partial \ell},
\end{eqnarray}
\begin{equation}
\label{5.5}
\partial_t\delta u_k+ay\frac{\partial}{\partial x}\delta u_k+a \delta_{kx}\delta u_y+
\frac{1}{mn_0} \left[(\partial_{n}P_{k\ell}^{(0)})\frac{\partial
\delta n}{\partial
r_\ell}+(\partial_{T}P_{k\ell}^{(0)})\frac{\partial \delta
T}{\partial r_\ell} -\eta_{k\ell i j} \frac{\partial^2 \delta
u_i}{\partial r_\ell r_j} \right]=0.
\end{equation}
Here, it is understood that the pressure tensor $P_{ij}^{(0)}$ and
its derivatives with respect to $n$ and $T$, the cooling rate
$\zeta_0$ and the transport coefficients $\eta_{ijk\ell}$,
$\mu_{ij}$, and $\kappa_{ij}$ are evaluated in the steady USF
state.

To analyze the linearized hydrodynamic equations (\ref{5.3})--(\ref{5.5})
it is convenient to transform to the local Lagrangian frame,
$r_i'=r_i-t a_{ij} r_j$. The Lees-Edwards boundary conditions then
become simple periodic boundary conditions in the variable ${\bf
r}'$ \cite{LD97}. A Fourier representation is defined as
\begin{equation}
\label{5.6} \delta \tilde{x}_{\mu} ({\bf k},t)=\int  d{\bf r}'\;
e^{i{\bf k}\cdot {\bf r}'}\delta x_{\mu}({\bf r},t)=\int  d{\bf
r}\; e^{i{\bf k}(t)\cdot {\bf r}}\delta x_{\mu}({\bf r},t),
\end{equation}
where in the second equality $k_i(t)=k_j(\delta_{ij}-t a_{ji})$.
Periodicity conditions requires that $k_i=2 n_i \pi/L_i$, where
$n_i$ are integers and $L_i$ are the linear dimensions of the
system. In this Fourier representation, the resulting set of five
linear equations defines the hydrodynamic modes, i.e., linear
response excitations to small perturbations. If at least one of
the modes grows in time, the reference USF state is linearly
unstable. Given the mathematical difficulties involved in the
general problem, for the sake of simplicity, here I consider two
kind of perturbations: (i) perturbations along the velocity
gradient direction only ($k_x=k_z=0; k_y\neq 0$) and (ii) perturbations in
the vorticity direction only ($k_x=k_y=0; k_z\neq 0$). In both cases, the
linearized hydrodynamic equations have time-independent
coefficients.

\subsection{Perturbations in the velocity gradient direction ($k_x=k_z=0; k_y\neq 0$)}

Let us consider first perturbations along the $y$ direction only.
In this case, Eqs.\ (\ref{5.3})--(\ref{5.5}) in this Fourier
representation can be written in the matrix form
\begin{equation}
\label{5.7}
\partial_{\tau} \delta \tilde{x}_\mu^*+F_{\mu \nu} \delta \tilde{x}_\nu^*=0,
\end{equation}
where the dimensionless quantities $\tau=\nu_0 t$ and $\delta
\tilde{x}_\mu^*\equiv \{\rho_k, \theta_k, {\bf w}_k\}$, with
\begin{equation}
\label{5.8} \rho_k=\frac{\delta \tilde{n}}{n_0}, \quad
\theta_k=\frac{\delta \tilde{T}}{T_0}, \quad {\bf
w}_k=\frac{\delta {\bf \tilde{u}}}{\sqrt{T_0/m}},
\end{equation}
have been introduced. The matrix $F_{\mu \nu}$ is
\begin{equation}
\label{5.9} F_{\mu \nu}=2 C \delta_{\mu 2}\delta_{\nu 1}+C
\delta_{\mu 2}\delta_{\nu 2}+a^*\delta_{\mu 3}\delta_{\nu 4}-i k^*
G_{\mu \nu}+k^{*2}H_{\mu \nu},
\end{equation}
where $a^*=a/\nu_0$, $\nu_0$ is the collision frequency
(\ref{2.20}) of the reference state and $k^*=\ell_0 k$,
$\ell_0=\sqrt{T_0/m}/\nu_0$ being of the order of the mean free
path. In addition, we have introduced the coefficient
\begin{equation}
\label{5.9.1} C(\alpha)=-\frac{1}{3}a^*\left(1+a^*\partial_a^*
\right)P_{xy}^*,
\end{equation}
and the square matrices
\begin{equation}
{\sf G}=\left(
\begin{array}{ccccc}
0 & 0 & 0&1&0 \\
0&0&\frac{2}{3}(P_{xy}^*-a^*\eta_{xyxy}^*)+\zeta_{xy}^*
&\frac{2}{3}(P_{yy}^*-a^*\eta_{xyyy}^*)+\zeta_{yy}^*&0\\
\left(1-a^*\partial_{a^*}\right)P_{xy}^*&\left(1-\frac{1}{2}a^*\partial_{a^*}\right)P_{xy}^*&0&0&0\\
\left(1-a^*\partial_{a^*}\right)P_{yy}^*&\left(1-\frac{1}{2}a^*\partial_{a^*}\right)P_{yy}^*&0&0&0\\
0&0&0&0&0
\end{array}
\right),
\label{5.10}
\end{equation}
\begin{equation}
{\sf H}=\left(
\begin{array}{ccccc}
0 & 0 & 0&0&0 \\
\frac{5}{3}\mu_{yy}^*&\frac{5}{3}\kappa_{yy}^*&0&0&0\\
0&0&\eta_{xyxy}^*&\eta_{xyyy}^*&0\\
0&0&\eta_{yyxy}^*&\eta_{yyyy}^*&0\\
0&0&0&0&\eta_{zyzy}^*
\end{array}
\right),
\label{5.11}
\end{equation}
have been also introduced. Here, $P_{ij}^*=P_{ij}^{(0)}/n_0T_0$ and
\begin{equation}
\label{5.12}
\zeta_{ij}^*=-\frac{1}{48}(1-\alpha^2)\left(P_{k\ell}^*-\delta_{k\ell}\right)\eta_{k\ell ij}^*.
\end{equation}

The eigenvalues $\lambda_\mu(k,\alpha)$  of the matrix ${\sf
F}(k,\alpha)$ determine the time evolution of $\delta
\tilde{x}_{\mu}^*(k,t)$. In the case that the real parts of the
eigenvalues $\lambda_{\mu}(k,\alpha)$ are positive, then the USF
state will be linearly stable. Before considering the general case
, it is convenient to consider some special limits. Thus, in the
elastic limit ($\alpha=1$), the hydrodynamic modes of the
Navier-Stokes equations (for the particular case considered here
and in the context of the BGK model) are recovered \cite{RL77},
namely, two sound modes, a heat mode and a two-fold degenerate
shear mode. To second order in $k^*$ they are given by
\begin{equation}
\label{5.14} \lambda_{\mu}(k,\alpha=1)\rightarrow
\left\{i\sqrt{\frac{5}{3}}k^*+k^{*2},
-i\sqrt{\frac{5}{3}}k^*+k^{*2}, k^{*2}, k^{*2}, k^{*2}\right\},
\end{equation}
and consequently, excitations around equilibrium are damped. It is
also quite illustrative to get the modes by setting $k=0$, namely,
consider small, homogenous deviations from the steady shear flow
state. In this case, it is easy to see that $\rho_k$ and $w_{y,k}$
are constant and
\begin{equation}
\label{5.14.1} w_{x,k}(\tau)=w_{x,k}(0)-a \tau w_{y,k}(0),
\end{equation}
\begin{equation}
\label{5.14.2} \theta_k(\tau)=\theta_k(0)e^{-C\tau}-2\rho_k(0).
\end{equation}
The mode associated with $w_{x,k}$ is unstable to an initial
perturbation in $w_{y,k}$, leading to an unbounded linear change
in time. However, stability is still possible at finite $k$ if
this behavior is modulated by exponential damping factors. With
respect to the temperature field, initial disturbances decay at
$\tau \to \infty$ if the coefficient $C(\alpha)>0$. Figure
\ref{fig5} shows that the coefficient $C$ is positive for any
value of $\alpha$ and so, this mode is stable with a finite decay
constant.

The analysis for $k\neq 0$ requires to get the eigenvalues
$\lambda_\mu(k^*,\alpha)$ with the full nonlinear dependence of
$k^*$. However, the structure of ${\sf F}(k,\alpha)$ shows that
the perturbation $\delta \tilde{x}_{5}^*\propto \delta
\tilde{u}_z$ is decoupled from the other four modes and hence can
be obtained more easily. This is due to the choice of gradients
along the $y$ direction only. The eigenvalue associated with this
mode is positive and is simply given by
\begin{equation}
\label{5.13}
\lambda_5(k,\alpha)=\eta_{zyzy}^*k^{*2},\quad \eta_{zyzy}^*=\frac{\beta}{(\beta+\zeta^*)^2},
\end{equation}
where $\zeta^*$ is defined by Eq.\ (\ref{2.21}). The remaining
modes correspond to $\rho_k$, $\theta_k$ and the components of the
velocity field $w_{x,k}$ and $w_{y,k}$. They are the solutions of
a quartic equation with coefficients that depend on $k^*$ and
$\alpha$. The results show that $\text{Re}\;
\lambda_{\mu}(k^*,\alpha)>0$ for all the values of the coefficient
of restitution $\alpha$ and consequently, the flow remains stable
to this kind of perturbations. As an illustration, the dispersion
relations for a gas with $\alpha=0.8$ are plotted in Fig.\
\ref{fig8}. It is apparent that all the  real parts of the
eigenvalues $\lambda_\mu$ are positive in the range of values of
wavenumber $k^*$ considered. Our conclusion agrees with previous
stability analysis \cite{S92,AN97} based on the Navier-Stokes
constitutive equations where it was found a minimum value of solid
fraction (around 0.156) below which the USF is stable. Given that
our system is a dilute gas (zero density), the present results
confirm previous findings when one uses the improved transport
coefficients.

\begin{figure}
\includegraphics[width=0.5 \columnwidth,angle=0]{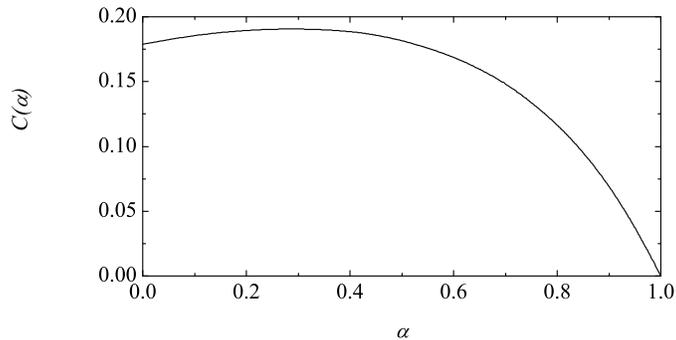}
\caption{Dependence of $C(\alpha)$ on the coefficient of
restitution $\alpha$. \label{fig5}}
\end{figure}

\begin{figure}
\includegraphics[width=0.5 \columnwidth,angle=0]{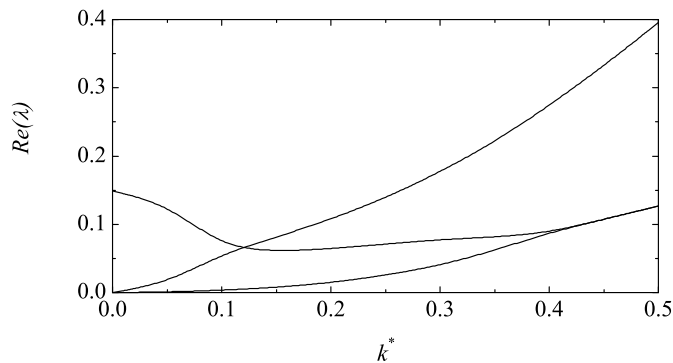}
\caption{Dispersion relations for a granular gas with
$\alpha=0.8$ in the case of perturbations along the velocity gradient direction.
Only the real parts of the eigenvalues is plotted.
\label{fig8}}
\end{figure}

\subsection{Perturbations in the vorticity direction ($k_x=k_y=0; k_z\neq 0$)}

The variation of the hydrodynamic modes with wavenumber $k=k_z$
in the vorticity direction is considered next. This situation has
not been widely studied in the literature since most of the
studies have been focussed on 2-D flows due to the relative
computational efficiency with which they can be analyzed. Here,
for the sake of simplicity, I consider perturbations for which
$\delta u_x=\delta u_y=0$ and so, the eigenvalues
$\lambda_\mu(k^*,\alpha)$ obey a cubic equation. The analysis is
similar to the one carried out in the previous section and so,
details will be omitted. For a given value of $\alpha$, it can be
seen that this dispersion relation has one real root and a complex
conjugate pair of damping modes. The instability arises from the
real root since this mode $ \lambda_{\mu}(k^*,\alpha)>0$ if $k^*$
is larger than a certain threshold value $k_s^*(\alpha)$. This
value can be obtained by solving $\lambda_{\mu}(k^*,\alpha)=0$. As
a consequence, the USF state is linearly stable against
excitations with a wavenumber $k^*>k_s^*(\alpha)$. The stability
line $k_s^*(\alpha)$ is plotted in Fig.\ \ref{fig4} as a function
of the coefficient of restitution. Above this line the modes are
stable, while below this line they are unstable.
For comparison, the corresponding stability line obtained
from the approximations made in previous works \cite{S92,AN97} is also
plotted. This line can be formally obtained from the results derived in this
paper when one replaces the expressions of the coefficients $\eta_{ijk\ell}$,
$\kappa_{ij}$, and $\mu_{ij}$ by their corresponding
Navier-Stokes expressions \cite{single}. It is apparent that the Navier-Stokes
approximation captures the qualitative dependence of $k_s^*$ on $\alpha$, although
as expected quantitative discrepancies between both descriptions appear
as the dissipation increases. Thus, for instance, for $\alpha=0.8$ the discrepancies
between both approaches are about 22 $\%$ while for $\alpha=0.5$ the discrepancies
are about 49$\%$. The prediction of
a long-wavelength instability for the USF state has been observed
in early molecular dynamics simulations \cite{MD} and
qualitatively agrees with the previous analytical results based on
the Navier-Stokes equations \cite{S92,B93,AN97,K00}. At a
quantitative level, the lack of numerical results from the
Boltzmann equation prevent us to carry out a more detailed
comparison to confirm the results derived from this kinetic model.
We hope that the results offered here will stimulate the
performance of such computer simulations.

\begin{figure}
\includegraphics[width=0.5 \columnwidth,angle=0]{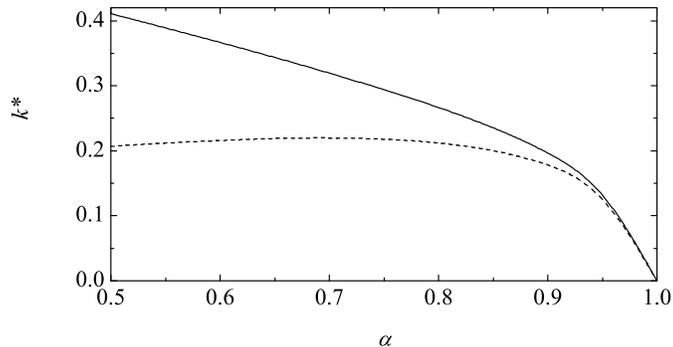}
\caption{Stability lines $k_s^*(\alpha)$ corresponding to the perturbation along the
vorticity direction. The solid line corresponds to the results derived here while the dashed
line refers to the results obtained from the Navier-Stokes approximation.
The region above the curve corresponds to the stable
domain, while the region below the curve corresponds to the
unstable domain. \label{fig4}}
\end{figure}

\section{Summary and Discussion}
\label{sec6}

The objective of this paper has been to study the transport
properties of a granular gas of inelastic hard spheres for the
special nonequilibrium states near the uniform (simple) shear flow
(USF). Although the derivation of the Navier-Stokes equations
(with explicit expressions for the transport coefficients
appearing in them) from a microscopic description has been widely
worked out in the past \cite{single,mixture}, the analysis of
transport in a strongly shearing granular gas has received little
attention due perhaps to its complexity and technical
difficulties. Very recently, a generalized Chapman-Enskog method
has been proposed to analyze transport around nonequilibrium
states in granular gases \cite{L05}. In the case of the USF state, due to the
anisotropy induced in the system by the presence of shear flow,
tensorial quantities are required to describe the momentum and
heat fluxes instead of the usual Navier-Stokes transport
coefficients \cite{single,mixture}. In this paper we have been
interested in a physical situation where {\em weak} spatial
gradients of density, velocity and temperature coexist with a {\em
strong} shear rate. Under these conditions, the corresponding
generalized transport coefficients characterizing heat and
momentum transport are {\em nonlinear} functions of both the
(reduced) shear rate $a^*$ and the coefficient of restitution
$\alpha$. The determination of such transport coefficients has
been the primary aim of this paper.

Due to the difficulties embodied in this problem, a low-density
gas described by the inelastic Boltzmann equation has been
considered. Although the exact solution to the Boltzmann equation
in the (steady) USF is not known, a good estimate of the relevant
transport properties can be obtained by means of Grad's method
\cite{MG02,G02,SGD04}. The reliability of this approximation has
been recently assessed by comparison with Monte Carlo simulations
of the Boltzmann equation \cite{MG02,AS05}. Assuming that the USF
state is slightly perturbed, the Boltzmann equation has been
solved by a Chapman-Enskog-like expansion where the shear flow
state is used as the reference state rather than the local
equilibrium or the (local) homogeneous cooling state. Due to the
spatial dependence of the zeroth-order distribution $f^{(0)}$
(reference state), this distribution is not in general stationary
and only in very special conditions has a simple relation with the
(steady) USF distribution \cite{L05}. Here, since one the main
goals has been to address a stability analysis of the USF state,
for practical purposes my results have been specialized to the
steady state, namely, when the hydrodynamic variables satisfy
the balance condition (\ref{3.12}).
In this situation, the (reduced) shear rate $a^*$
is coupled with the coefficient of restitution $\alpha$ [see Eq.\
(\ref{2.24})] so that the latter is the relevant parameter of the
problem. In the first order of the expansion the momentum and heat
fluxes are given by Eqs.\ (\ref{3.18}) and (\ref{3.19}),
respectively, where the set of generalized transport coefficients
$\eta_{ijk\ell}$, $\mu_{ij}$, and $\kappa_{ij}$ are given in terms
of the solutions of the linear integral equations
(\ref{3.15bis})--(\ref{3.17bis}). As expected, there are many new
transport coefficients in comparison to the case of states near
equilibrium or cooling state. These coefficients provide all the
information on the physical mechanisms involved in the transport
of momentum and energy under shear flow.

Practical applications require to solve the integral equations
(\ref{3.15bis})--(\ref{3.17bis}), which is in general quite a
complex problem. In addition, the fourth-degree velocity moments
of USF (whose evaluation would require to consider higher-order
terms in Grad's solution (\ref{2.17}) of the Boltzmann equation)
are needed to determine the coefficients $\kappa_{ij}$ and
$\mu_{ij}$. To overcome such mathematical difficulties, here a
kinetic model of the Boltzmann equation \cite{BDS99} has been
used. This kinetic model can be considered as an extension of the
well-known BGK equation to inelastic gases. Although the kinetic
model is only a crude representation of the Boltzmann equation, it
does preserve the most important features for transport, such as
the homogeneous cooling state and the macroscopic conservation
laws. The model has a free parameter $\beta$ to be adjusted to fit
a given property of the Boltzmann equation. Here, $\beta$ is given
by Eq.\ (\ref{2.22}) to get good quantitative agreement of the
Navier-Stokes shear viscosity coefficient obtained from the
Boltzmann equation. Furthermore, this choice yields the same
results for rheological properties in the USF problem as those
derived from the Boltzmann equation by means of Grad's method. On
the other hand, given that the model does not intend to mimic the
behavior of the true distribution function beyond the thermal
velocity region, discrepancies between the kinetic model and the
Boltzmann equation are expected beyond the second-degree velocity
moments (which quantify the elements of the pressure tensor).
Nevertheless, a recent comparison with Monte Carlo simulations of
the Boltzmann equation \cite{AS05} have shown the accuracy of the
kinetic model predictions for the fourth-degree moments. As
illustrated in Fig.\ \ref{fig6}, the semi-quantitative agreement
between theory and simulation is not restricted to the
quasielastic limit ($\alpha \thickapprox 0.99$) since it covers
values of large dissipation ($\alpha \gtrsim 0.5$). The use of
this kinetic model allows one to get the explicit dependence of
the generalized transport coefficients on the coefficient of
restitution. This dependence has been illustrated in some cases
showing that in general the deviation of the transport
coefficients from their corresponding elastic values is quite
significant.

With these new expressions for the fluxes, a closed set of
generalized hydrodynamic equations for states close to USF has
been derived. A stability analysis of these linearized
hydrodynamic equations with respect to the USF state have been
also carried out to identify the conditions for stability in terms
of dissipation. Two different kind of perturbations to the USF
state has been analyzed: (i) perturbations along the velocity
gradient only ($k_y\neq 0$) and (ii) perturbations along the
vorticity direction only ($k_z\neq 0$). In the first case, previous
results \cite{S92,AN97} have shown that the USF is stable for a
dilute gas while the USF becomes unstable in the second case for
all $\alpha$ \cite{L04}. These results agree with these findings and
the USF is unstable for any finite value of dissipation at
sufficiently long wave lengths when disturbances are generated in
the orthogonal direction to the shear flow plane. On the other hand, as
expected, quantitative discrepancies between our results and those given
\cite{S92,AN97} from the Navier-Stokes approximation become significant as
the dissipation increases. These differences have been illustrated in
Fig.\ \ref{fig4} for the stability line.
Although the
instability of the USF has been extensively studied for many
authors by using a Navier-Stokes description \cite{S92,B93,AN97}
as well as solutions of the Boltzmann equation in the quasielastic
limit \cite{K00,K01}, I am not aware of any previous solution of
the hydrodynamic equations where the generalized transport
coefficients describing transport around USF were taken into
account. The analytical results found in this paper allows a
quantitative comparison with numerical solutions to the Boltzmann
equation for finite dissipation. As happens for the USF problem
for elastic \cite{LD97,LDMSL96,MSLDL98} and inelastic
\cite{MG02,AS05} gases, one expects that the results reported here
compare well with such simulations, confirming again the
reliability of the kinetic theory results to characterize the
onset and the first stages of evolution of the clustering
instability. We hope to carry out these simulations in the next
future.

On the other hand, the stability analysis performed here has only
considered spatial variations along the $y$ and $z$ directions.
More complex dynamics is expected in the general case of arbitrary
direction for the spatial perturbation. This will be worked
elsewhere along with comparison with direct Monte Carlo computer
simulations of the Boltzmann equation. Another possible direction
of study is the extension of the present approach to other
physically interesting reference states, such as the nonlinear
Couette flow. This is a more realistic shearing problem than the
USF state since combined heat and momentum transport appears in
the system. Given that an exact solution to the kinetic model used
here is known for the Couette flow problem \cite{TTMGSD01}, the
reference distribution for the Chapman-Enskog-like expansion is
available.

\acknowledgments
I am grateful to Dr. James Lutsko for pointing
out on some conceptual errors that were present in a previous
version of this manuscript. It is also a pleasure to thank to
Dr. Andr\'es Santos for valuable discussions.
Partial support of the Ministerio de
Ciencia y Tecnolog\'{\i}a (Spain) through Grant No. FIS2004-01399
(partially financed by FEDER funds) is acknowledged.

\appendix
\section{Chapman-Enskog expansion}
\label{appA}

Inserting the expansions (\ref{3.7}) and (\ref{3.9}) into Eq.\
(\ref{3.1}), one gets the kinetic equation for $f^{(1)}$,
\begin{equation}
\label{a1}
\left(\partial_t^{(0)}-aV_y\frac{\partial}{\partial V_x}+{\cal L}\right)f^{(1)}=
-\left[\partial_t^{(1)}+({\bf V}+{\bf u}_0)\cdot \nabla \right]f^{(0)},
\end{equation}
where ${\cal L}$ is the linearized Boltzmann collision operator
\begin{equation}
\label{a2} {\cal L}X \equiv
-\left(J[f^{(0)},X]+J[X,f^{(0)}]\right).
\end{equation}
The velocity dependence on the right side of Eq.\ (\ref{a1}) can be obtained from the macroscopic
balance equations to first order in the gradients. They are given by
\begin{equation}
\label{a3}
\partial_t^{(1)}n+{\bf u}_0\cdot \nabla n=-\nabla \cdot (n\delta {\bf u}),
\end{equation}
\begin{equation}
\label{a4}
\partial_t^{(1)}\delta {\bf u}+({\bf u}_0+\delta {\bf u})\cdot \nabla \delta {\bf u}=
-\frac{1}{\rho}\nabla \cdot {\sf P}^{(0)},
\end{equation}
\begin{equation}
\label{a5}
\frac{3}{2}n\partial_t^{(1)}T+\frac{3}{2}n({\bf u}_0+
\delta {\bf u})\cdot \nabla T+aP_{xy}^{(1)}+{\sf P}^{(0)}:\nabla
\delta {\bf u}=-\frac{3}{2}p\zeta^{(1)},
\end{equation}
where $\rho=mn$ is the mass density,
\begin{equation}
\label{a6}
P_{ij}^{(1)}=\int d{\bf c}\, m c_i c_j  f^{(1)}({\bf c}),
\end{equation}
and
\begin{equation}
\label{a7} \zeta^{(1)}=\frac{1}{3p}\int d{\bf c}\,mc^2{\cal L}
f^{(1)}.
\end{equation}
Use of Eqs.\ (\ref{a3})--(\ref{a5}) in Eq.\ (\ref{a1}) yields
\begin{equation}
\label{a8}
\left(\partial_t^{(0)}-aV_y\frac{\partial}{\partial V_x}+{\cal L}\right)
f^{(1)}-\zeta^{(1)}T\frac{\partial f^{(0)}}{\partial T}={\bf Y}_n\cdot \nabla n+{\bf Y}_T\cdot \nabla T
+{\sf Y}_u:\nabla \delta {\bf u},
\end{equation}
where
\begin{equation}
\label{a9}
Y_{n,i}=-\frac{\partial f^{(0)}}{\partial n}c_i+\frac{1}{\rho}
\frac{\partial f^{(0)}}{\partial \delta u_j}\frac{\partial P_{ij}^{(0)}}{\partial n},
\end{equation}
\begin{equation}
\label{a10}
Y_{T,i}=-\frac{\partial f^{(0)}}{\partial T}c_i+\frac{1}{\rho}
\frac{\partial f^{(0)}}{\partial \delta u_j}\frac{\partial P_{ij}^{(0)}}{\partial T},
\end{equation}
\begin{equation}
\label{a11}
Y_{u,ij}=n\frac{\partial f^{(0)}}{\partial n}\delta_{ij}-\frac{\partial f^{(0)}}
{\partial \delta u_i}c_j+\frac{2}{3n}\frac{\partial f^{(0)}}{\partial T}\left(P_{ij}^{(0)}-a\eta_{xyij}\right).
\end{equation}
According to Eqs.\ (\ref{a9})--(\ref{a10}), $Y_{u,ij}$ has the
same symmetry properties (\ref{2.15.1}) as the distribution
function $f^{(0)}$ while $Y_{n,i}$ and $Y_{T,i}$ are odd functions
in the velocity ${\bf c}$.

The solution to Eq.\ (\ref{a8}) has the form
\begin{equation}
\label{a12}
f^{(1)}=X_{n,i}({\bf c})\nabla_i n+ X_{T,i}({\bf c})\nabla_i T+X_{u,ji}({\bf c})\nabla_i \delta u_j.
\end{equation}
Note that in Eq.\ (\ref{a11}) the coefficients $\eta_{ijk\ell}$ are defined through Eq.\ (\ref{3.20}).
Substitution of the solution (\ref{a12}) into the relation (\ref{a7}) allows one to write the cooling
rate in the form
\begin{equation}
\label{a12.1}
\zeta^{(1)}=\zeta_{n,i} \nabla_i n+ \zeta_{T,i}\nabla_i T
+\zeta_{u,ji}\nabla_i \delta u_j,
\end{equation}
where
\begin{equation}
\label{a12.2}
\left(
\begin{array}{c}
\zeta_{n,i}\\
\zeta_{T,i}\\
\zeta_{u,ij}
\end{array}
\right)=
\frac{1}{3p}\int\, d{\bf c}\, mc^2 {\cal L}
\left(
\begin{array}{c}
X_{n,i}\\
X_{T,i}\\
X_{u,ij}
\end{array}
\right).
\end{equation}
However, given that $X_{n,i}$ and $X_{T,i}$ are odd functions in
${\bf c}$ [see for instance, Eqs.\ (\ref{a15}) and (\ref{a16}) below],
the terms proportional to $\nabla n$ and $\nabla T$  vanish
by symmetry, i.e.,
\begin{equation}
\label{a13.1}
\zeta_{n,i}=\zeta_{T,i}=0.
\end{equation}
Thus, the only nonzero contribution to $\zeta^{(1)}$
comes from the term proportional to the tensor $\nabla_i\delta u_j$:
\begin{equation}
\label{a13}
\zeta^{(1)}={\zeta}_{u,ji}\nabla_{i} \delta u_j.
\end{equation}
An estimate of the tensor $\zeta_{u,ij}$
has been made in Appendix \ref{appC} by considering the leading
terms in a Sonine polynomial expansion of the distribution
$f^{(1)}$. Its expression is given by Eq.\ (\ref{c10}). As expected, $\zeta_{u,ij}$ vanishes
in the elastic limit ($\alpha=1$).

The coefficients $X_{n,i}$, $X_{T,i}$, and $X_{u,ij}$ are functions of the peculiar velocity ${\bf c}$ and
the hydrodynamic fields. In addition, there are contributions from the time derivative $\partial_t^{(0)}$
acting on the temperature and velocity gradients given by
\begin{eqnarray}
\label{a13.1}
\partial_t^{(0)} \nabla_i T&=&\nabla_i\partial_t^{(0)}T
\nonumber\\
&=&\left(\frac{2a}{3n^2}(1-n\partial_n)
P_{xy}^{(0)}-\frac{\zeta^{(0)}T}{n}\right) \nabla_i n-\left(
\frac{2a}{3n}\partial_T
P_{xy}^{(0)}+\frac{3}{2}\zeta^{(0)}\right)\nabla_i T,
\end{eqnarray}
\begin{equation}
\label{a13.2}
\partial_t^{(0)} \nabla_i \delta u_j=\nabla_i \partial_t^{(0)} \delta u_j=-a_{jk} \nabla_i \delta u_k.
\end{equation}
Substituting (\ref{a13}) into (\ref{a8}) and identifying coefficients
of independent gradients gives the set of equations
\begin{equation}
\label{a15}
-\left[\left(\frac{2}{3n}a
P_{xy}^{(0)}+T\zeta^{(0)}\right)\partial_T+ a
c_y\frac{\partial}{\partial c_x}-{\cal
L}\right]X_{n,i}+\frac{T}{n}\left[\frac{2a}{3p}(1-n\partial_n)
P_{xy}^{(0)}-\zeta^{(0)}\right]X_{T,i}=Y_{n,i},
\end{equation}
\begin{equation}
\label{a16} -\left[\left(\frac{2}{3n}a
P_{xy}^{(0)}+T\zeta^{(0)}\right)\partial_T+
\frac{2a}{3p}T(\partial_T P_{xy}^{(0)})+\frac{3}{2}\zeta^{(0)}+a
c_y\frac{\partial}{\partial c_x}-{\cal L}\right]X_{T,i}=Y_{T,i},
\end{equation}
\begin{equation}
\label{a17}
-\left[\left(\frac{2}{3n}a
P_{xy}^{(0)}+T\zeta^{(0)}\right)\partial_T+a
c_y\frac{\partial}{\partial c_x}-{\cal L}\right]X_{u,k\ell}-
a\delta_{ky}X_{u,x\ell}-\zeta_{u,k\ell}T\partial_T
f^{(0)}=Y_{u,k\ell}.
\end{equation}
Upon writing Eqs.\ (\ref{a15})--(\ref{a17}), use has been made of the property
\begin{eqnarray}
\label{a18}
\partial_t^{(0)} X &=&\frac{\partial X}{\partial T}\partial_t^{(0)}
T+\frac{\partial X}{\partial \delta u_i}\partial_t^{(0)}
\delta u_i \nonumber\\
&=&-\left(\frac{2}{3n}a
P_{xy}^{(0)}+T\zeta^{(0)}\right)\frac{\partial X}{\partial T}
+a_{ij}\delta u_j \frac{\partial X}{\partial c_i},
\end{eqnarray}
where in the last step we have taken into account that $X$ depends on
$\delta {\bf u}$ through ${\bf c}={\bf V}-\delta {\bf u}$.

\section{Evaluation of the cooling rate}
\label{appC}

In this Appendix the contribution $\zeta_{u,ij}$ to the
cooling rate $\zeta^{(1)}$ is evaluated by expanding
$X_{u,ij}$ as series in Sonine
polynomials and taking the lowest order truncation.
The tensor $\zeta_{u,ij}$ is given by
\begin{eqnarray}
\label{c5}
\zeta_{u,ij}&=&\frac{1}{3p}\int\, d{\bf c}_1\, mc_1^2 {\cal L}X_{u,ij}\nonumber\\
&=&-\frac{1}{3p}\int\, d{\bf c}_1\, mc_1^2\left\{J[{\bf c}_1|f^{(0)},X_{u,ij}]
+J[{\bf c}_1|X_{u,ij},f^{(0)}]\right\}.
\end{eqnarray}
A useful identity for an arbitrary function $h({\bf c}_1)$ is
\begin{equation}
\label{c6}
\int d{\bf c}_1h({\bf c}_1)J[{\bf c}_1|f,g]=\sigma^{2}
\int d{\bf c}_1\int d{\bf c}_2
f({\bf c}_1)g({\bf c}_2)\int\,
d\widehat{\boldsymbol {\sigma}}\,\Theta (\widehat{\boldsymbol {\sigma}}\cdot
{\bf g})(\widehat{\boldsymbol {\sigma}}\cdot {\bf g})
\left[h({\bf c}_1'')-h({\bf c}_1)\right],
\end{equation}
where ${\bf g}={\bf c}_1-{\bf c}_2$ and
\begin{equation}
\label{c7}
{\bf c}_1''={\bf c}_1-\frac{1}{2}(1+\alpha)
(\widehat{\boldsymbol {\sigma}}\cdot {\bf g})\widehat{\boldsymbol {\sigma}}.
\end{equation}
Using (\ref{c6}), Eq.\ (\ref{c5}) can be written as
\begin{equation}
\label{c8}
\zeta_{u,ij}=\frac{m}{6p}\sigma^{2}(1-\alpha^2)
\int d{\bf c}_1\int d{\bf c}_2\;
f^{(0)}({\bf c}_1)X_{u,ij}({\bf c}_2)
\int d\widehat{\boldsymbol {\sigma}}\,\Theta
(\widehat{\boldsymbol {\sigma}}\cdot
{\bf g})(\widehat{\boldsymbol {\sigma}}\cdot {\bf g})^3.
\end{equation}
The integration over $\widehat{\boldsymbol {\sigma}}$ in (\ref{c8}) yields
\begin{equation}
\label{c9}
\zeta_{u,ij}=\frac{m}{12p}\pi \sigma^{2}(1-\alpha^2)
\int d{\bf c}_1\int d{\bf c}_2\;f^{(0)}({\bf c}_1)X_{u,ij}({\bf c}_2).
\end{equation}
This equation is still exact. To perform the integrals over ${\bf c}_1$ and ${\bf c}_2$
one takes the Grad approximation (\ref{2.17}) to $f^{(0)}$ and expands $X_{u,ij}$ in
Sonine polynomials. In this case and
according to the anisotropy of the USF problem, one takes the approximation
\begin{equation}
\label{c2}
X_{u,k\ell}({\bf c}) \to -\frac{1}{2nT^2}D_{ij}\eta_{ijk\ell}f_{0}(c),
\end{equation}
where
\begin{equation}
\label{c3}
f_{0}(c)=n\left(\frac{m}{2\pi T}\right)^{3/2}\exp\left(-\frac{mc^2}{2T}\right)
\end{equation}
is the Maxwellian distribution and
\begin{equation}
\label{c4}
D_{ij}({\bf c})=m\left(c_i c_j-\frac{1}{3}c^2\delta_{ij}\right).
\end{equation}
Next, change variables to the (dimensionless) relative velocity
${\bf g}^*=({\bf c}_1-{\bf c}_2)/v_0$ and center of mass ${\bf G}^*=({\bf c}_1+{\bf c}_2)/2v_0$,
where $v_0=\sqrt{2T/m}$ is the thermal velocity.
A lengthy calculation leads to
\begin{eqnarray}
\label{c10}
\zeta_{u,ij}&=&-\frac{1}{6}\frac{v_0\sigma^2}{\pi^2 T}(1-\alpha^2)
\int d{\bf g}^*\int d{\bf G}^* \;g^{*3} e^{-2G^{*2}} e^{-g^{*2}/2} \nonumber\\
& & \times \left[G_k^*G_{\ell}^*G_m^*G_n^*-\frac{1}{18}g^{*2}G^{*2}\left(\delta_{km}\delta_{\ell n}+
\delta_{kn}\delta_{\ell m}\right)+\frac{1}{16}g_k^* g_{\ell}^*g_m^* g_n^*\right]
\left(\frac{P_{mn}}{nT}-\delta_{mn}\right)\eta_{k\ell ij}\nonumber\\
&=&-\frac{1}{15}\sigma^2\sqrt{\frac{\pi}{m T}}(1-\alpha^2)
\left(\frac{P_{k\ell}}{nT}-\delta_{k\ell}\right)\eta_{k\ell ij}.
\end{eqnarray}
Of course, when $\alpha=1$, then $\zeta_{u,ij}=0$.

\section{Behavior of the zeroth-order velocity moments near the steady state}
\label{appD}

This Appendix addresses the behavior of the velocity moments of
the zeroth-order distribution $f^{(0)}$ near the steady state. Let
us start with the elements of the pressure tensor $P_{ij}^{(0)}$.
In the context of the Boltzmann equation and by using Grad's
approximation (\ref{2.17}), they verify the equation
\begin{equation}
\label{d1} -\left(\frac{2}{3n}a
P_{xy}^{(0)}+T\zeta^{(0)}\right)\frac{\partial}{\partial T}
P_{ij}^{(0)}+ a_{i\ell}P_{j\ell}^{(0)}+a_{j\ell}P_{i\ell}^{(0)}=
-\nu\left[\beta
\left(P_{ij}^{(0)}-p\delta_{ij}\right)+\zeta^*P_{ij}^{(0)}\right].
\end{equation}
Since we are interested in the {\em hydrodynamic} solution, the temperature derivative term can be
written as
\begin{equation}
\label{d2}
T\partial_T P_{ij}^{(0)}=T\partial_T p P_{ij}^*=p\left(1-\frac{1}{2}a^*\frac{\partial}
{\partial a^*}\right)P_{ij}^*,
\end{equation}
where $P_{ij}^*=P_{ij}^{(0)}/p$. Upon deriving (\ref{d2}), use has been made of the fact that
the dimensionless pressure tensor $P_{ij}^*$ depends on $T$ only through
its dependence on the reduced shear rate
$a^*=a/\nu(n, T)$. In dimensionless form, the set of equations (\ref{d1}) become
\begin{equation}
\label{d3}
-\left(\frac{2}{3}a^*
P_{xy}^{*}+\zeta^{*}\right)\left(1-\frac{1}{2}a^*\frac{\partial}
{\partial a^*}\right)P_{ij}^*+
a_{i\ell}^*P_{j\ell}^{*}+a_{j\ell}^*P_{i\ell}^{*}=
-\left[\beta
\left(P_{ij}^{*}-\delta_{ij}\right)+\zeta^*P_{ij}^{*}\right],
\end{equation}
where
\begin{equation}
\label{d4}
\zeta^*=\frac{\zeta^{(0)}}{\nu}=\frac{5}{12}(1-\alpha^2).
\end{equation}
Let us consider the elements $P_{xy}^*$ and $P_{yy}^*=P_{zz}^*$.
From Eq.\ (\ref{d1}), one gets
\begin{equation}
\label{d5} -\left(\frac{2}{3}a^*
P_{xy}^{*}+\zeta^{*}\right)\left(1-\frac{1}{2}a^*\frac{\partial}
{\partial a^*}\right)P_{xy}^* + a^*P_{yy}^{*}=
-\left(\beta+\zeta^* \right)P_{xy}^{*},
\end{equation}
\begin{equation}
\label{d6} -\left(\frac{2}{3}a^*
P_{xy}^{*}+\zeta^{*}\right)\left(1-\frac{1}{2}a^*\frac{\partial}
{\partial a^*}\right)P_{yy}^*= -\left(\beta+\zeta^*
\right)P_{yy}^{*}+\beta.
\end{equation}
This set of equations have a singular point corresponding to the
steady state solution, i.e., when $a^*(T)=a^*_s$ where
$a^*_s(\alpha)$ is the steady state value of $a^*$ given by Eq.\
(\ref{2.24}). Since we are interested in the solution of Eqs.\
(\ref{d5}) and (\ref{d6}) near the steady state, we assume that in
this region $P_{xy}^*$ and $P_{yy}^*$ behave as
\begin{equation}
\label{d7}
P_{xy}^*=P_{xy,s}^*+\left(\frac{\partial P_{xy}^*}{\partial a^*}\right)_s(a^*-a^*_s)+\cdots,
\end{equation}
\begin{equation}
\label{d8}
P_{yy}^*=P_{yy,s}^*+\left(\frac{\partial P_{yy}^*}{\partial a^*}\right)_s(a^*-a^*_s)+\cdots,
\end{equation}
where the subscript $s$ means that the quantities are evaluated in the steady state. Substitution of
(\ref{d7}) and (\ref{d8}) into Eqs.\ (\ref{d5}) and (\ref{d6}) allows one to determine the corresponding
derivatives. The result is
\begin{equation}
\label{d9}
\left(\frac{\partial P_{yy}^*}{\partial a^*}\right)_s=4 P_{yy,s}^*\frac{a^*_s C+P_{xy,s}^*}{2a^{*2}_s C
+6 \beta+3\zeta^*},
\end{equation}
where $C\equiv \left(\partial P_{xy}^*/\partial a^*\right)_s$ is
the real root of the cubic equation
\begin{equation}
\label{d10} 2 a_s^{*4}
C^3+12a_s^{*2}(\zeta^*+\beta)C^2+\frac{9}{2}(7\zeta^{*2}
+14\zeta^*\beta+4\beta^2)C+9\beta(\zeta^*+\beta)^{-2}(
2\beta^2-2\zeta^{*2}-\beta\zeta^*).
\end{equation}
Equations (\ref{d9}) and (\ref{d10}) can be also obtained from a
different way. Let us write the set of equations (\ref{d5}) and
(\ref{d6}) as
\begin{equation}
\label{d11}
\frac{\partial P_{xy}^*}{\partial a^*}=\frac{-2P_{yy}^*-\frac{2}{a^*}P_{xy}^*\left( \beta-\frac{2}{3}
P_{xy}^*a^*\right)}{\zeta^*+\frac{2}{3}a^*P_{xy}^*},
\end{equation}
\begin{equation}
\label{d12}
\frac{\partial P_{yy}^*}{\partial a^*}=\frac{2\beta-2P_{yy}^*\left( \beta-\frac{2}{3}
P_{xy}^*a^*\right)}{a^*\left(\zeta^*+\frac{2}{3}a^*P_{xy}^*\right)}.
\end{equation}
In the steady state limit ($a^*\to a_s^*$), the numerators and
denominators of Eqs.\ (\ref{d11}) and (\ref{d12}) vanish.
Evaluating the corresponding limit by means of l'Hopital's rule,
one reobtains the above results (\ref{d9}) and (\ref{d10}). This
procedure can be used to get the behavior of the remaining
velocity moments near the steady state.

The behavior of the fourth-degree velocity moments of the distribution $f^{(0)}$ near the steady
state is also needed to determine the transport coefficients $\mu_{ij}$ and $\kappa_{ij}$ associated
with the heat flux in the first-order solution.
To evaluate this behavior we use the Boltzmann kinetic model (\ref{4.1}).
Let us introduce the velocity moments of the zeroth-order distribution
\begin{equation}
\label{d13} M_{k_1,k_2,k_3}^{(0)}=\int \; d{\bf c}\;
c_x^{k_1}c_y^{k_2}c_z^{k_3} f^{(0)}({\bf c})
\end{equation}
These moments verify the equation
\begin{eqnarray}
\label{d14}
-\left(\frac{2}{3n}a
P_{xy}^{(0)}+T\zeta^{(0)}\right)\partial_T M_{k_1,k_2,k_3}^{(0)}&+&
a k_1 M_{k_1-1,k_2+1,k_3}^{(0)}=
-\nu \beta \left( M_{k_1,k_2,k_3}^{(0)}\right.\nonumber\\
& &\left.-N_{k_1,k_2,k_3}\right)-k\frac{\zeta^{(0)}}{2}
M_{k_1,k_2,k_3}^{(0)},
\end{eqnarray}
where $k=k_1+k_2+k_3$ and $N_{k_1,k_2,k_3}$ are the velocity
moments of the Gaussian distribution $f_0$. As before, the
derivative $\partial_T M_{k_1,k_2,k_3}^{(0)}$ can be written as
\begin{eqnarray}
\label{d15}
T\partial_T M_{k_1,k_2,k_3}^{(0)}&=&T\partial_T n\left(\frac{2T}{m}
\right)^{k/2}M_{k_1,k_2,k_3}^*(a^*)\nonumber\\
&=&n\left(\frac{2T}{m}\right)^{k/2} \frac{1}{2}\left(k
-a^*\partial_{a^*}\right)M_{k_1,k_2,k_3}^{*}(a^*).
\end{eqnarray}
In dimensionless form, Eq.\ (\ref{d14}) become
\begin{eqnarray}
\label{d16} -\left(\frac{2}{3}a^*
P_{xy}^{*}+\zeta^*\right)\frac{1}{2}(k
-a^*\partial_{a^*})M_{k_1,k_2,k_3}^{*}&+& k_1 a^*
M_{k_1-1,k_2+1,k_3}^{*}+
(\beta+\frac{k}{2}\zeta^*)M_{k_1,k_2,k_3}^{*}
\nonumber\\
& &
-\beta N_{k_1,k_2,k_3}^{*}=0,
\end{eqnarray}
where $N_{k_1,k_2,k_3}^{*}$ are the reduced moments of the the Gaussian
distribution given by
\begin{equation}
\label{d17}
N_{k_1,k_2,k_3}^{*}=\pi^{-3/2}\Gamma \left(\frac{k_1+1}{2}\right)
\left(\frac{k_2+1}{2}\right)\left(\frac{k_3+1}{2}\right)
\end{equation}
if $k_1$, $k_2$, and $k_3$ are even, being zero otherwise.
Equation (\ref{d16}) gives the expressions of the reduced moments
$M_{k_1,k_2,k_3,s}^{*}$ in the steady state. To get
$\partial_{a^*} M_{k_1,k_2,k_3}^{*}$ in the steady state, we
differentiate with respect to $a^*$ both sides of Eq.\ (\ref{d16})
and then takes the limit $a\to a_s^*$. In general, it is easy to
see that the problem becomes linear so that it can be easily
solved. To illustrate the procedure, let us consider for
simplicity the moment $M_{040}^*$, which obeys the equation
\begin{equation}
\label{d18} -\left(\frac{2}{3}a^* P_{xy}^{*}+\zeta^*\right)\left(2
-\frac{1}{2}a^*\partial_{a^*}\right)M_{040}^{*}
+\left(\beta+2\zeta^*\right)M_{040}^{*}-\frac{3}{4}\beta =0.
\end{equation}
From this equation, one gets the identity
\begin{eqnarray}
\label{d19} -\left(\frac{2}{3}a^*
P_{xy}^{*}+\zeta^*\right)\partial_{a^*}\left[\left( 2
-\frac{1}{2}a^*\partial_{a^*}\right)
M_{040}^{*}\right]&-&\frac{2}{3}\left[ P_{xy}^{*}+a^*
(\partial_{a^*}P_{xy}^*)\right]
\left(2 -\frac{1}{2}a^*\partial_{a^*}\right)M_{040}^{*}\nonumber\\
& & +(\beta+2\zeta^*)\partial_{a^*} M_{040}^{*}=0
\end{eqnarray}
In the steady state limit, Eq.\ (\ref{3.12}) applies and the first
term on the left hand side vanishes. In this case, one easily gets
\begin{equation}
\label{d20}
\left(\frac{\partial}{\partial a^*} M_{040}^*\right)_s=\frac{4 \chi_s}
{a_s^* \chi_s+2 \beta+4 \zeta^*}M_{040,s}^*,
\end{equation}
where
\begin{equation}
\label{d21}
\chi_s=\frac{2}{3}\left[P_{xy,s}^{*}+a_s^* \left(\frac{\partial P_{xy}^*}{\partial a^*}\right)_s\right]
\end{equation}
is a known function and
\begin{equation}
\label{d22}
M_{040,s}^*=\frac{3}{4}\frac{\beta}{\beta+2\zeta^*}.
\end{equation}
Proceeding in a similar way, all the derivatives of the form
$\partial_{a^*} M^*$ can be analytically computed in the steady
state.

\section{Kinetic model results in the steady state}
\label{appB}

In this Appendix, I display the results obtained from the model kinetic
equation chosen here for the determination of the generalized transport coefficients.
In the model, the Boltzmann collision operator is replaced by the term \cite{BDS99}
\begin{equation}
\label{b1}
J[f,f]\to -\beta \nu (f-f_0)+\frac{\zeta}{2}\frac{\partial}{\partial
{\bf c}}\cdot \left({\bf c}f\right),
\end{equation}
where $\nu$ and $\beta$ are given by Eqs.\ (\ref{2.20}) and (\ref{2.22}), respectively,
$f_0$ is the local equilibrium distribution (\ref{2.18}) and $\zeta$ is
the cooling rate (\ref{2.12}).

\subsection{Steady state solution for the (unperturbed) USF}

Let us consider first the steady state solution to the (unperturbed) USF problem.
In this case, $\delta {\bf u}={\bf 0}$ and so ${\bf c}={\bf V}$.
The one-particle distribution function $f({\bf V})$ obeys the kinetic equation
\begin{equation}
\label{b2}
-aV_y\frac{\partial}{\partial V_x} f({\bf V})=-\beta \nu (f-f_0)+
\frac{\zeta^{(0)}}{2}\frac{\partial}{\partial {\bf V}}\cdot \left({\bf V}f\right) \;,
\end{equation}
where here $\zeta$ has been approximated by its local equilibrium approximation $\zeta^{(0)}$
given by Eq.\ (\ref{2.21}). The main advantage of using a kinetic model instead of the
Boltzmann equation is that the model lends itself to an exact solution \cite{GS03,AS05}.
It can be written as
\begin{equation}
\label{b3}
f({\bf V})=n\left(\frac{m}{2T}\right)^{3/2} f^*({\boldsymbol \xi}), \quad
{\boldsymbol \xi}=\sqrt{\frac{m}{2T}}{\bf V},
\end{equation}
where the reduced velocity distribution function $f^*$ is a function of the coefficient
of restitution $\alpha$ and the reduced peculiar velocity ${\boldsymbol \xi}$:
\begin{equation}
\label{b4}
f^*({\boldsymbol \xi})=\pi^{-3/2}\int_0^{\infty}\;ds \;e^{-(1-\frac{3}{2}\overline{\zeta})s}
\exp\left[-e^{\overline{\zeta}s}\left({\boldsymbol \xi}+s\overline{{\sf a}}\cdot {\boldsymbol \xi}
\right)^2\right].
\end{equation}
Here, $\overline{{\sf a}}={\sf a}/(\nu \beta)$ and $\overline{\zeta}=\zeta^{(0)}/(\nu \beta)$.
It has been recently shown that the distribution function (\ref{b4}) presents an excellent
agreement with Monte Carlo simulations in the region of thermal velocities, even for strong
dissipation \cite{AS05}.

The explicitly knowledge of
the velocity distribution function allows one to compute {\em all} the velocity moments. We introduce
the moments
\begin{equation}
\label{b5}
M_{k_1,k_2,k_3}=\int \; d{\bf v}\; V_x^{k_1}V_y^{k_2}V_z^{k_3} f({\bf V})
\end{equation}
According to the symmetry of the USF distribution (\ref{b4}), the only nonvanishing moments correspond to
even values of $k_1+k_2$ and $k_3$. In this case, after some algebra, one gets \cite{AS05}
\begin{equation}
\label{b5.1}
M_{k_1,k_2,k_3}=
n\left(\frac{2T}{m}\right)^{k/2}M_{k_1,k_2,k_3}^*,
\end{equation}
where the reduced moments $M_{k_1,k_2,k_3}^*$ are given by
\begin{eqnarray}
\label{b6}
M_{k_1,k_2,k_3}^*&=&
\pi^{-3/2}
\sum_{\stackrel{q=0}{q+k_1=\text{even}}}^{k_1}
(-\overline{a})^q\left(1+\frac{\overline{\zeta}}{2}k\right)^{-(1+q)}
\frac{k_1!}{(k_1-q)!}
\nonumber\\
& & \times
\;\Gamma\left(\frac{k_1-q+1}{2}\right)\Gamma\left(\frac{k_2+q+1}{2}\right)
\Gamma\left(\frac{k_3+1}{2}\right),
\end{eqnarray}
with $\overline{a}=a/(\nu \beta)=a^*/\beta$. It is easy to see that
the expressions for the second degree-degree
velocity moments (rheological properties) coincide with those given from
the Boltzmann equation by using Grad's approximation, Eqs.\ (\ref{2.23})--(\ref{2.24}).

\subsection{Transport coefficients}

Let us now evaluate the generalized transport coefficients $\eta_{ijk\ell}, \kappa_{ij}$, and
$\mu_{ij}$ in the steady state.
They can be obtained from Eqs.\ (\ref{3.15bis})--(\ref{3.17bis}) with the replacement
given by Eqs.\ (\ref{4.2.1}) and (\ref{4.2.2}).
With these changes, Eqs.\ (\ref{3.15bis})--(\ref{3.17bis}) become
\begin{equation}
\label{b8} \left(-a c_y\frac{\partial}{\partial c_x}+ \nu
\beta-\frac{\zeta^{(0)}}{2}\frac{\partial}{\partial {\bf c}}\cdot
{\bf c}\right)X_{n,i}
+\frac{2a}{3}\frac{T}{n}(P_{xy}^*+a^*\partial_{a^*}
P_{xy}^{*})X_{T,i}=Y_{n,i},
\end{equation}
\begin{equation}
\label{b9}  \left(-a c_y\frac{\partial}{\partial
c_x}-\frac{1}{3}a\left(
 P_{xy}^*-a^*\partial_{a^*}P_{xy}^*\right)+{\cal L}\right)X_{T,i}=Y_{T,i},
\end{equation}
\begin{equation}
\label{b10} \left(-a c_y\frac{\partial}{\partial c_x}+\nu \beta -
\frac{\zeta^{(0)}}{2}\frac{\partial}{\partial {\bf c}}\cdot {\bf
c}\right)X_{u,j\ell}
-\frac{1}{2}\zeta_{u,j\ell}\left[\frac{\partial}{\partial {\bf
c}}\cdot ({\bf c}f^{(0)}) +2T\frac{\partial}{\partial
T}f^{(0)}\right] -a\delta_{jy}X_{u,x\ell}=Y_{u,j\ell}.
\end{equation}

In order to get the transport coefficients $\kappa_{ij}$, $\mu_{ij}$, and $\eta_{ijk\ell}$,
it is convenient to introduce the velocity moments
\begin{equation}
\label{b11}
A_{k_1,k_2,k_3}^{(i)}=\int\, d{\bf c}\, c_x^{k_1}c_y^{k_2}c_z^{k_3}X_{n,i},
\end{equation}
\begin{equation}
\label{b12}
B_{k_1,k_2,k_3}^{(i)}=\int\, d{\bf c}\, c_x^{k_1}c_y^{k_2}c_z^{k_3}X_{T,i},
\end{equation}
\begin{equation}
\label{b13}
C_{k_1,k_2,k_3}^{(ij)}=\int\, d{\bf c}\, c_x^{k_1}c_y^{k_2}c_z^{k_3}X_{u,ij}.
\end{equation}
The knowledge of these moments allows one to get all the
transport coefficients of the {\em perturbed} USF problem.
Now, we multiply Eqs.\ (\ref{b8})--(\ref{b10}) by $c_x^{k_1}c_y^{k_2}c_z^{k_3}$
and integrate over velocity. The result is
\begin{equation}
\label{b14}
ak_1A_{k_1-1,k_2+1,k_3}^{(i)}+\left(\nu \beta+\frac{1}{2}k\zeta^{(0)}\right)
A_{k_1,k_2,k_3}^{(i)}+\omega_n B_{k_1,k_2,k_3}^{(i)}=
\int\, d{\bf c}\, c_x^{k_1}c_y^{k_2}c_z^{k_3}Y_{n,i},
\end{equation}
\begin{equation}
\label{b15}
ak_1B_{k_1-1,k_2+1,k_3}^{(i)}+\left(\nu \beta+\frac{1}{2}k\zeta^{(0)}+\omega_T\right)
B_{k_1,k_2,k_3}^{(i)}=
\int\, d{\bf c}\, c_x^{k_1}c_y^{k_2}c_z^{k_3}Y_{T,i},
\end{equation}
\begin{eqnarray}
\label{b16} ak_1C_{k_1-1,k_2+1,k_3}^{(j\ell)}+\left(\nu
\beta+\frac{1}{2}k\zeta^{(0)} \right)& &
C_{k_1,k_2,k_3}^{(j\ell)}+\frac{1}{2}\zeta_{u,j\ell}\left(k-2T\partial_T\right)
M_{k_1,k_2,k_3}^{(0)}\nonumber\\
& &
-a\delta_{jy}C_{k_1,k_2,k_3}^{(x\ell)}=
\int\, d{\bf c}\, c_x^{k_1}c_y^{k_2}c_z^{k_3}Y_{u,j\ell}.
\end{eqnarray}
Here, $M_{k_1,k_2,k_3}^{(0)}$ are the moments of the zeroth-order distribution $f^{(0)}$ and we have
introduced the quantities
\begin{equation}
\label{b16.1}
\omega_n=\frac{2a}{3}\frac{T}{n}(P_{xy}^*+a^*\partial_{a^*}
P_{xy}^{*}), \quad \omega_T=-\frac{1}{3}a\left(
 P_{xy}^*-a^*\partial_{a^*}P_{xy}^*\right).
\end{equation}
The right-hand side terms of Eqs.\ (\ref{b14})--(\ref{b16}) can be easily evaluated with the result
\begin{eqnarray}
\label{b17}
{\cal A}_{k_1,k_2,k_3}^{(\ell)}&\equiv& \int\, d{\bf c}\, c_x^{k_1}c_y^{k_2}c_z^{k_3}Y_{n,\ell}\nonumber\\
&=&-\frac{\partial}{\partial n}M_{k_1+\delta_{\ell x},k_2+\delta_{\ell y},k_3+\delta_{\ell z}}+\frac{1}{\rho}
\frac{\partial P_{\ell j}^{(0)}}{\partial n}\nonumber\\
& & \times \left(\delta_{jx}k_1M_{k_1-1,k_2,k_3}+\delta_{jy}k_2M_{k_1,k_2-1,k_3}+\delta_{jz}k_3M_{k_1,k_2,k_3-1}\right)
\nonumber\\
&=&-\left(\frac{2T}{m}\right)^{\frac{k+1}{2}}\left[
(1-a^*\partial_{a^*})M_{k_1+\delta_{\ell x},k_2+\delta_{\ell
y},k_3+\delta_{\ell z}}^*-
\frac{1}{2}(1-a^*\partial_{a^*})P_{\ell j}^{*}\right.\nonumber\\
& & \left. \times
 \left(\delta_{jx}k_1M_{k_1-1,k_2,k_3}^*+\delta_{jy}k_2M_{k_1,k_2-1,k_3}^*+
 \delta_{jz}k_3M_{k_1,k_2,k_3-1}^*\right)\right],
 \end{eqnarray}
\begin{eqnarray}
\label{b18}
{\cal B}_{k_1,k_2,k_3}^{(\ell)}&\equiv& \int\, d{\bf c}\, c_x^{k_1}c_y^{k_2}c_z^{k_3}Y_{T,\ell}\nonumber\\
&=& -\frac{\partial}{\partial T}M_{k_1+\delta_{\ell x},k_2+\delta_{\ell y},k_3+\delta_{\ell z}}+\frac{1}{\rho}
\frac{\partial P_{\ell j}^{(0)}}{\partial T}\nonumber\\
& & \times \left(\delta_{jx}k_1M_{k_1-1,k_2,k_3}+\delta_{jy}k_2M_{k_1,k_2-1,k_3}+\delta_{jz}k_3M_{k_1,k_2,k_3-1}\right)
\nonumber\\
&=&-n\left(\frac{2T}{m}\right)^{\frac{k+1}{2}}\left[\frac{1}{2T}
(k+1-a^*\partial_{a^*})M_{k_1+\delta_{\ell x},k_2+\delta_{\ell
y},k_3+\delta_{\ell z}}^*-
\frac{1}{2T}(1-\frac{1}{2}a^*\partial_{a^*})P_{\ell j}^{*}\right.\nonumber\\
& & \left. \times
\left(\delta_{jx}k_1M_{k_1-1,k_2,k_3}^*+\delta_{jy}k_2M_{k_1,k_2-1,k_3}^*+
\delta_{jz}k_3M_{k_1,k_2,k_3-1}^*\right)\right],
\end{eqnarray}
\begin{eqnarray}
\label{b19}
{\cal C}_{k_1,k_2,k_3}^{(j\ell)}&\equiv& \int\, d{\bf c}\, c_x^{k_1}c_y^{k_2}c_z^{k_3}Y_{u,j\ell}\nonumber\\
&=&-\delta_{j\ell}\left(1-n\frac{\partial}{\partial n}\right)M_{k_1,k_2,k_3}+\frac{2}{3n}
\left(P_{j\ell}^{(0)}-a\eta_{xyj\ell}\right)
\frac{\partial}{\partial T}M_{k_1,k_2,k_3}\nonumber\\
& & -M_{k_1,k_2,k_3}\left(\delta_{jx}\delta_{\ell x}k_1+\delta_{jy}
\delta_{\ell y}k_2+\delta_{jz}\delta_{\ell z}k_3\right)\nonumber\\
& & -k_1\delta_{jx}\left(\delta_{\ell y}M_{k_1-1,k_2+1,k_3}+
\delta_{\ell z}M_{k_1-1,k_2,k_3+1}\right)\nonumber\\
& & -k_2\delta_{jy}\left(\delta_{\ell x}M_{k_1+1,k_2-1,k_3}+
\delta_{\ell z}M_{k_1,k_2-1,k_3+1}\right)\nonumber\\
& & -k_3\delta_{jz}\left(\delta_{\ell x}M_{k_1+1,k_2,k_3-1}+
\delta_{\ell y}M_{k_1,k_2+1,k_3-1}\right)\nonumber\\
&=&-n\left(\frac{2T}{m}\right)^{k/2}\left[
\delta_{j\ell} a^*\partial_{a^*} M_{k_1,k_2,k_3}^*\right.\nonumber\\
& & -
\frac{1}{3n
T}\left(P_{j\ell}^{(0)}-a\eta_{xyj\ell}\right)(k-a^*\partial_{a^*})M_{k_1,k_2,k_3}^*
\nonumber\\
& & +M_{k_1,k_2,k_3}^*\left(\delta_{jx}\delta_{\ell x}k_1+
\delta_{jy}\delta_{\ell y}k_2+\delta_{jz}\delta_{\ell z}k_3\right)\nonumber\\
& & +k_1\delta_{jx}\left(\delta_{\ell y}M_{k_1-1,k_2+1,k_3}^*+
\delta_{\ell z}M_{k_1-1,k_2,k_3+1}^*\right)\nonumber\\
& & +k_2\delta_{jy}\left(\delta_{\ell x}M_{k_1+1,k_2-1,k_3}^*+
\delta_{\ell z}M_{k_1,k_2-1,k_3+1}^*\right)\nonumber\\
& & \left.+k_3\delta_{jz}\left(\delta_{\ell x}M_{k_1+1,k_2,k_3-1}^*+
\delta_{\ell y}M_{k_1,k_2+1,k_3-1}^*\right)\right].
\end{eqnarray}
Here, $M_{k_1,k_2,k_3}^*$ are the reduced moments of the
distribution $f^{(0)}$ defined by Eq.\ (\ref{b5.1}). In the steady
state, $M_{k_1,k_2,k_3}^*$ is given by Eq.\ (\ref{b6}) while the
derivatives $\partial_{a^*} M_{k_1,k_2,k_3}^*$ can be obtained by
following the procedure described in Appendix \ref{appD}.

The solution to Eqs.\ (\ref{b14})--(\ref{b16}) can be written as
\begin{equation}
\label{b20}
A_{k_1,k_2,k_3}^{(i)}=(\nu \beta)^{-1}\sum_{q=0}^{k_1}
(-\overline{a})^q\left(1+\frac{k\overline{\zeta}}{2}\right)^{-(1+q)}\frac{k_1!}{(k_1-q)!}\left[
{\cal A}_{k_1-q,k_2+q,k_3}^{(i)}-\omega_n B_{k_1-q,k_2+q,k_3}^{(i)}\right],
\end{equation}
\begin{equation}
\label{b21}
B_{k_1,k_2,k_3}^{(i)}=(\nu \beta)^{-1}\sum_{q=0}^{k_1}
(-\overline{a})^q\left(1+\overline{\omega}_T
+\frac{k\overline{\zeta}}{2}\right)^{-(1+q)}\frac{k_1!}{(k_1-q)!}
{\cal B}_{k_1-q,k_2+q,k_3}^{(i)},
\end{equation}
\begin{eqnarray}
\label{b22}
C_{k_1,k_2,k_3}^{(j\ell)}&=&(\nu \beta)^{-1}\sum_{q=0}^{k_1}
(-\overline{a})^q
\left(1+\frac{k\overline{\zeta}}{2}\right)^{-(1+q)}\frac{k_1!}{(k_1-q)!}\nonumber\\
& & \times \left[ {\cal
C}_{k_1-q,k_2+q,k_3}^{(j\ell)}+a\delta_{jy}C_{k_1-q,k_2+q,k_3}^{(x\ell)}
-\frac{1}{2}n\left(\frac{2T}{m}\right)^{k/2}\zeta_{u,j\ell}\;a^*\partial_{a^*}
M_{k_1-q,k_2+q,k_3}^*\right], \nonumber\\
\end{eqnarray}
where $\overline{\omega}_T=\omega_T/(\nu \beta)$. From Eqs.\
(\ref{b20})--(\ref{b22}) one can get the expressions for the
transport coefficients $\kappa_{ij}$, $\mu_{ij}$, and
$\eta_{ijk\ell}$ in terms of $\beta$, $\overline{\zeta}$ and
$\overline{a}$.


\begin{thebibliography} {99}


\bibitem{GS95}A. Goldshtein and M. Shapiro, J. Fluid Mech. {\bf 282}, 41 (1995).

\bibitem{BDS97}J. J. Brey, J. W. Dufty, and A. Santos, J. Stat. Phys. {\bf 87}, 1051 (1997).

\bibitem{CC70}S. Chapman and T. G. Cowling, {\em The Mathematical Theory of Nonuniform Gases} (Cambridge University Press, Cambridge, 1970).

\bibitem{single}J. J. Brey, J. W. Dufty, C. S. Kim, and A. Santos, Phys. Rev. E {\bf 58}, 4638 (1998); J. J. Brey and D. Cubero, in {\em Granular Gases}, Lecture Notes in Physics, edited by T. P{\"o}schel and S. Luding (Springer Verlag, Berlin, 2001), p. 59; V. Garz\'o and J. M. Montanero, Physica A {\bf 313}, 336 (2002).

\bibitem{mixture} V. Garz\'o and J. W. Dufty, Phys. Fluids {\bf 14}, 1476 (2002);
V. Garz\'o and J. M. Montanero, Phys. Rev. E {\bf 69}, 021301 (2004).

\bibitem{DSMC}J. J. Brey, M. J. Ruiz-Montero, and D. Cubero, Europhys. Lett. {\bf 48}, 359 (1999);
V. Garz\'o and J. M. Montanero, Physica A {\bf 313}, 336 (2002); J. M. Montanero and V. Garz\'o,
Phys. Rev. E {\bf 67}, 021308 (2003).

\bibitem{GS03}V. Garz\'o and A. Santos, {\em Kinetic Theory of Gases in Shear Flows. Nonlinear Transport}
(Kluwer Academic, Dordrecht, 2003).

\bibitem{L05}J. Lutsko, cond-mat/0510749.

\bibitem{C90}See, for instance, C. S. Campbell, Annu. Rev. Fluid Mech. {\bf 22}, 57 (1990);
I. Goldhirsch, Annu. Rev. Fluid Mech. {\bf 35}, 267 (12003).

\bibitem{LSJC84}C. K. K. Lun, S. B. Savage, D. J. Jeffrey, and N. Chepurniy, J. Fluid Mech.
{\bf 140}, 223 (1984).

\bibitem{JR88}J. T. Jenkins and M. W. Richman, J. Fluid Mech. {\bf 192}, 313 (1988).

\bibitem{C89}C. S. Campbell, J. Fluid Mech. {\bf 203}, 449 (1989).

\bibitem{HS92}M. A. Hopkins and  H. H. Shen, J. Fluid Mech. {\bf 244}, 477 (1992).

\bibitem{SK94}P. J. Schmid and H. K. Kyt\"omaa, J. Fluid Mech. {\bf 264}, 255 (1994).

\bibitem{LB94}C. K. K. Lun and A. A. Bent, J. Fluid Mech. {\bf 258}, 335 (1994).

\bibitem{GT96}I. Goldhirsch and M. L. Tan, Phys. Fluids {\bf 8}, 1752 (1996).

\bibitem{SGN96}N. Sela, I. Goldhirsch, and S. H. Noskowicz, Phys. Fluids {\bf 8}, 2337 (1997).

\bibitem{BRM97}J. J. Brey, M. J. Ruiz-Montero, and F. Moreno, Phys. Rev. E {\bf 55}, 2846 (1997).

\bibitem{CR98}C.-S. Chou and M. W. Richman, Physica A {\bf 259}, 430 (1998); C.-S. Chou {\em ibid.}
{\bf 287}, 127 (2000); {\bf 290}, 341 (2001).

\bibitem{MGSB99}J. M. Montanero, V. Garz\'o, A. Santos, and J. J. Brey, J. Fluid Mech. {\bf 389}, 391 (1999).

\bibitem{CH02}R. Clelland and C. M. Hrenya, Phys. Rev. E {\bf 65}, 031301 (2002).

\bibitem{MG02}J. M. Montanero and V. Garz\'o, Physica A {\bf 310}, 17 (2002);
Mol. Sim. {\bf 29}, 357 (2003); V. Garz\'o and J. M. Montanero, Granular Matter {\bf 5}, 165 (2003).

\bibitem{G02}V. Garz\'o, Phys. Rev. E {\bf 66}, 021308 (2002).

\bibitem{L04}J. Lutsko, Phys. Rev. E {\bf 70}, 061101 (2004).

\bibitem{AL03}M. Alam and S. Luding, J. Fluid Mech. {\bf 476}, 69 (2003).

\bibitem{MGAL05}J. M. Montanero, V. Garz\'o, M. Alam, and S. Luding, cond-mat/0411548.

\bibitem{SGD04}A. Santos, V. Garz\'o, and J. W. Dufty, Phys. Rev. E {\bf 69}, 061303 (2004).

\bibitem{AS05}A. Santos and A. Astillero, Phys. Rev. E {\bf 72},
031308 (2005); A. Astillero and A. Santos, Phys. Rev. E {\bf 72},
031309 (2005).

\bibitem{LD97}M. Lee and J. W. Dufty, Phys. Rev. E {\bf 56}, 1733 (1997).

\bibitem{TTMGSD01}M. Tij, E. Tahiri, J. M. Montanero, V. Garz\'o, A. Santos, and J. W. Dufty,
J. Stat. Phys. {\bf 103}, 1035 (2001).

\bibitem{BDS99}J. J. Brey, J. W. Dufty, and A. Santos, J. Stat. Phys. {\bf 97}, 281 (1999).

\bibitem{MD}O. R. Walton and R. L. Braun, J. Rheol. {\bf 30}, 949 (1986); Acta Mech. {\bf 73},
86 (1986); C. S. Campbell and C. E. Brennen, J. Fluid Mech. {\bf 151}, 167 (1985); M. H. Hopkins and
M. Y. Louge, Phys. Fluids A {\bf 3}, 47 (1991); M. Alam and S. Luding, Phys. Fluids {\bf 15}, 2298
(2003).

\bibitem{S92}S. B. Savage, J. Fluid Mech. {\bf 241}, 109 (1992).

\bibitem{B93}M. Babic, J. Fluid Mech. {\bf 254}, 127 (1993).

\bibitem{AN97}M. Alam and P. R. Nott, J. Fluid Mech. {\bf 343}, 267 (1997); {\bf 377}, 99
(1998); J. D. Goddard and M. Alam, Particulate Science and
Technology {\bf 17}, 69 (1999).

\bibitem{K00}V. Kumaran, Physica A {\bf 275}, 283 (2000); {\bf 284}, 246 (2000).

\bibitem{K01}V. Kumaran, Phys. Fluids {\bf 13}, 2258 (2001).


\bibitem{LE72}A. W. Lees and S. F. Edwards, J. Phys. C {\bf 5}, 1921 (1972).

\bibitem{DSBR86}J. W. Dufty, A. Santos, J. J. Brey, and R. F. Rodr\'{\i}guez, Phys. Rev. A {\bf 33}, 459 (1986).


\bibitem{note}The model kinetic equation of Ref.\ \cite{BDS99} uses the exact local homogeneous
cooling state of the Boltzmann equation for the distribution $f_0$. An excellent approximation
to $f_0$ in the region of thermal velocities is given by
the local equilibrium distribution function (\ref{2.18}). Given that here we are mainly interested
in computing transport properties, the local equilibrium form (\ref{2.18})
is adopted for $f_0$.


\bibitem{RL77}P. Resibois and M. de Leener, {\em Classical Kinetic Theory of Fluids} (Wiley, New York, 1977).


\bibitem{L04}J. F. Lutsko, cond-mat/0403551.

\bibitem{LDMSL96}M. Lee, J. W. Dufty, J. M. Montanero, A. Santos, and J. F. Lutsko, Phys. Rev. Lett.
{\bf 76}, 2702 (1996).



\bibitem{MSLDL98}J. M. Montanero, A. Santos, M. Lee, J. W. Dufty, and J. F. Lutsko,
Phys. Rev. E {\bf 57}, 546 (1998).




\end{thebibliography}
\end{document}